\documentclass[12pt,a4paper]{article}
\usepackage{amsmath}
\usepackage{amsfonts}
\usepackage{amssymb}
\usepackage{setspace}
\usepackage{amsthm}
\usepackage{rotating}
\usepackage[a4paper, left=2.5cm, right=2.5cm, top=2.8cm, bottom=2.8cm]{geometry}
\usepackage{float}

\begin{document}

\begin{center}
\textbf{\large On Control Charts for Monitoring the Variance of a Time Series}\\[0.5cm]
\textsc{Taras Lazariv, Wolfgang Schmid, and Svitlana Zabolotska  }\\[1cm]
{\it \footnotesize  Department of Statistics, European University Viadrina, PO Box 1786, D-15207 Frankfurt (Oder), Germany}\\
\end{center}

\begin{abstract}
\onehalfspacing
In this paper we derive control charts for the variance of a Gaussian process using the likelihood ratio approach, the generalized likelihood ratio approach, the sequential probability ratio method and a generalized sequential probability ratio procedure, the Shiryaev-Roberts procedure and a generalized Shiryaev-Roberts approach. Recursive presentations for the calculation of the control statistics are given for autoregressive processes of order 1. In an extensive simulation study these schemes are compared with existing control charts for the variance. In order to asses the performance of the schemes both the average run length and the average delay are used.
\end{abstract}
\section*{}
\emph{Keywords}: control charts; CUSUM charts; generalized likelihood ratio; SPRT; Shiryaev-Roberts procedures; variance changes; time series; statistical process control; sequential detection

\section{Introduction}
\onehalfspacing
In many applications we are faced with the problem to detect changes over time in an observed process. Because usually a change should be detected
fast sequential methods are more appropriate in such a situation. The most important tools for monitoring a process are control charts (cf. Stoumbos et al. (2000)). Control charts are successfully applied in engineering for a long time (e.g., Lawson and Kleinman (2005), Fris\'{e}n (2007)). In the last 20 years many further applications in different areas have been studied like, e.g., in public health, economics, environmental sciences.  In that context the underlying processes have a more complicate structure and are mostly modeled by a time series. Alwan and Roberts (1988) showed that control charts for independent variables cannot be directly applied to time series. They proposed to use residual charts, i.e. to transform the original observations such that the transformed observations are independent and to apply the well-known control charts to these residuals. Residual charts have been studied by several authors (e.g., Harris and Ross (1991), Wardell et al. (1994), Lu and Reynolds (1999)). Another possibility is to directly monitor the observed process. The behavior of the Shewhart chart for time series was studied in Schmid (1995). An extension of the exponentially weighted moving average (EWMA) chart of Roberts (1959) to time series was proposed by Schmid (1997a). Cumulative sum (CUSUM) charts for time-dependent processes have been studied among others by Nikiforov (1975), Schmid (1997b), Fris\'{e}n and Knoth (2012). An overview on control charts for time series is given in Knoth and Schmid (2004) and Okhrin and Schmid (2007).

Most of the literature on that topic is dealing with the monitoring of the mean behavior of the observed process. Here we want to focus on the surveillance of the variance of a time series. We are interested to detect an increase in the variance. For instance, such a question is of great importance in economics
where the variance is the most applied measure for the risk and an early detection of a change in the risk behavior of an asset is an important information for an analyst. The first EWMA control chart for time series was introduced by MacGregor and Harris (1993). Schipper and Schmid (2001) introduced several one-sided variance charts for stationary processes, however, their main focus was in the area of nonlinear time series. In many applications the control statistic for independent processes is used and the independent process is replaced by the time series. Thus the structure of the time series is not taken into account for the derivation of the control statistic. This is a disadvantage of this procedure.

In this paper we derive control charts for the variance of a Gaussian process by making use of the likelihood ratio approach (LR), the sequential probability ratio test of Wald (SPRT), and the Shiryaev-Roberts (SR) procedure. For deriving these charts it is first assumed that the size of the change is known. Thus all obtained charts depend on a reference value which has to be suitably chosen in advance. This is sometimes a drawback in applications. We consider generalized control charts as well. They are obtained via the generalized LR, SPRT, and SR procedure. The great advantage of these schemes is that they do not depend on a reference value. It has to be emphasized that our results are quite general and cover all autoregressive moving average processes with a Gaussian white noise.

In Section 2 the underlying model of the paper is introduced. It is explained how in our paper the target process and the observed process are related with each other. In Section 3 the CUSUM control chart for the variance in the independent case is briefly presented and a new CUSUM control scheme for Gaussian processes is derived over the LR approach. In an example we consider the special cases of autoregressive processes of order $1$ and $2$ and it is shown that in that case the control statistics can be calculated recursively. In Section 4 a CUSUM variance chart is derived by the SPRT and the result is a residual chart. In Section 5 the Shiryaev-Roberts method is used to get a control chart for the variance. In the Sections 6 to 8 generalized control schemes are derived. In Section 6 we use the generalized LR method, in Section 7 a generalization of the SPRT approach, and in Section 8 a generalization of a modified version of the SR method is obtained.

In an extensive simulation study these control schemes are compared with each other assuming that the underlying target process is an autoregressive process of order 1 (AR(1)) (Section 9). As a measure for the performance of a control scheme the average run length (ARL) and the average delay are taken. All charts are calibrated such that the in-control ARL is the same if no change is present. Our results show that except the SR chart all other schemes with a reference value have the smallest out-of-control ARL if the reference value is equal to the true value of the change. It turns out that the generalized SR chart has the smallest ARL if the change is small. It is even better than the charts with the optimal reference value. For medium and larger changes the LR chart and the SPRT chart provides better results provided that the reference value is not dramatically smaller than the true change. Except the generalized SPRT scheme for all charts the worst average delay is equal to the average run length. The limit of the average delay seems to be the smallest for the GSR chart. This scheme must be preferred for larger changes if the change arises at a later time point.

\section{Modelling}
The aim of statistical process control (SPC) consists in detecting structural deviations within a process over time. It is examined whether the present observations can be considered as realizations of a given target process $\{Y_t\}$. The procedure is a sequential one. The observations (samples) are analyzed consecutively. It is desirable to detect a change as quickly as possible after its occurrence. Of course there are various types of changes which may influence the target process. In this paper we focus on the detection of an increase in the variance. Such a problem arises in practice very often. For instance, the variance is considered as a risk measure in economics and thus an increasing variance is a hint that the risk of an asset is getting larger. In engineering the variance reflects the quality of a production process and an increase is a bad sign since the production is getting worse.

Let $\{Y_t\}$ be a (weakly) stationary process with mean $\mu$ and autocovariance function $Cov(Y_t, Y_{t+h})=\gamma(h)$. In what follows it is assumed that the relationship between the target process $\{Y_t\}$ and the observed process $\{X_t\}$ is given by
\begin{equation}\label{model}
X_t=
\begin{cases}
	Y_t        & \text{for} \quad     1 \leq t < \tau\\
	\mu + \Delta (Y_t - \mu) & \text{for} \quad     t \geq \tau
\end{cases}
\end{equation}
for $t \in \mathbb{Z}$ with $\Delta > 1$ and $\tau \in \mathbb{N} \cup \{\infty\}$. Thus a change in the scale appears at position $\tau$ if $\tau < \infty$.
The observed process is said to be out of control. Else, if $\tau = \infty$, then $\{X_t\}$ is called to be in control. Here it is assumed that at a given time point exactly one observation is available.

Note that the change in the scale does not influence the mean structure. It holds that $E(X_t)=\mu$ as well in the in-control state as in the out-of-control state. Moreover, we get that

\begin{align*}
Var(X_t)&=
\begin{cases}
	Var(Y_t)        & \text{for} \quad     1 \leq t < \tau\\
	\Delta^2 Var(Y_t) & \text{for} \quad     t \geq \tau
\end{cases},
\\
Cov(X_t, X_{t+h})&=
\begin{cases}
	\gamma_h          & \text{for} \quad     t < \min\{\tau,\tau-h\}\\
	\Delta \gamma_h   & \text{for} \quad     \min\{\tau,\tau-h\} \leq t < \max\{\tau,\tau-h\}\\
	\Delta^2 \gamma_h & \text{for} \quad     t \geq \max\{\tau,\tau-h\}
\end{cases} .
\end{align*}
Thus the observed process $\{X_t\}$ is not stationary in the out-of-control case.

\section{A Variance Chart based on the Likelihood Ratio Approach}
\subsection{LR Approach applied to Independent Variables}

The CUSUM chart for the mean was introduced by Page (1954). The CUSUM scheme is a control chart with memory. At each time point the decision is based not only on the present observation, but also on previous observations. The CUSUM chart gives all former observations the same weight. Moreover, the control chart depends on an additional design parameter, the reference value. Up to now mainly CUSUM charts for the mean have been considered in literature.

CUSUM charts can be derived by means of the log likelihood ratio approach (see Siegmund (1985, Ch. II.6)). In that context it is demanded that $\Delta$ is a known quantity. For instance, assuming that the variables ${Y_t}$ are independent and normally distributed with expectation $\mu$ and variance $\gamma_0$ the log likelihood ratio of the present model is given by (see, e.g., Hawkins and Olwell (1998) and Schipper and Schmid (2001))

\[
\frac{1-1/\Delta^2}{2} ( \tilde{S}_n(\Delta) - \tilde{S}_{\tau-1}(\Delta) )
\]
for $n\geq\tau$ with
\begin{equation}\label{Summe}
\tilde{S}_n(\Delta) = \sum_{i=1}^{n} \frac{(X_i-\mu)^2}{\gamma_0} - n K(\Delta)
\end{equation}
and
\begin{equation}\label{Referenz}
K(\Delta)=\frac{\log(\Delta^2)}{1-1/\Delta^{2}} .
\end{equation}
We conclude that the process is out of control at time $n \ge 1$ if
\begin{equation}\label{Splus}
	S_n^+(\Delta) = \tilde{S}_n(\Delta)  - \min_{0 \leq i \leq n} \tilde{S}_i(\Delta) > c  .
\end{equation}
Note that $S_n^+(\Delta)$ can be calculated recursively as $S_n^+(\Delta) = \max \{0, S_{n-1}^+(\Delta) + (X_n-\mu)^2/\gamma_0 - K(\Delta)\}$
for $n\geq1$ with $S_0^+(\Delta) = 0$. This representation dramatically simplifies the practical calculation of the control design.

Unfortunately $\Delta$ is unknown in practice and for that reason it is necessary to fix a value for $\Delta$, say $\Delta^*$. In practice $\Delta^* > 1$ is interpreted as the change against which we want to be protected. Then we have to replace $K(\Delta)$ by $K(\Delta^*)$. Thus we obtain
\begin{equation}\label{Rekursion}
S_n^+(\Delta^*) = \max\{0, S_{n-1}^+(\Delta^*) + \frac{(X_n - \mu)^2}{\gamma_0} - K(\Delta^*)\},  n\geq1, S_0^+(\Delta^*) = 0 .
\end{equation}

The process is concluded to be out of control at time $n \ge 1$ if $S_n^+(\Delta^*) > c$. For each fixed value of $\Delta^*$, we determine the value of c such that the in-control ARL is equal to some predetermined quantity $\xi$.

In the above derivation it is assumed that the underlying process is independent. In this paper, however, we are interested in target processes which follow a time series. Because of the complicated structure of the likelihood function of a time series practitioners have used the above recursion (\ref{Rekursion}) for time series as well (e.g., Schipper and Schmid (2001)). Then $\{X_t\}$ stands for a time series, $\mu$ is equal to its mean and $\gamma_0$ is the in-control variance. It has to be noted that the control limit $c$ must be chosen by taking the time series structure into account. The run length of this scheme is given by
\begin{equation}\label{Niid}
N_{LR,iid}(c; \Delta^*) = \inf\{ n \in \mathbb{N}: S_n^+(\Delta^*) > c \}
\end{equation}
with $S_n^+(\Delta^*)$ as in (\ref{Rekursion}).

While for independent observations several numerical methods are available to calculate the ARL (cf. Brook  and Evans (1972), Knoth (2010)) the determination of this quantity turns out to be very difficult if the target process has a dependence structure. In that case no explicit formulas for the ARL are available and in practice the control limits are determined via simulations.

\subsection{LR Approach applied to Gaussian Processes}

In this section $\{Y_t\}$ is assumed to be a Gaussian process with mean zero and covariance function $k(i,j) = Cov(X_i, X_j)$. Assume that the covariance matrix $[ \kappa(i,j) ]_{i,j=1,..,n}$ is non-singular for each $n \ge 1$. Let $\hat{Y}_t$ denote the best linear predictor of $Y_t$ based on $Y_{t-1},..,Y_1$. This quantity can be recursively calculated using the innovations algorithm (cf. Brockwell and Davis (1991, p.172))
\[ \hat{Y}_t = \left\{ \begin{array}{ccc}
0 & \mbox{for} & t=1\\
\sum_{j=1}^{t-1} \theta_{tj} (Y_{t-j} - \hat{Y}_{t-j}) & \mbox{for} & t \ge 2
\end{array} \right. . \]
The quantities $\theta_{tj}$ can be determined recursively. Note that $\hat{Y}_t = \sum_{j=1}^{t-1} a_{tj} Y_j$ with some coefficients $a_{tj}$. Note that  for a Gaussian process the best linear predictor is equal to the predictor obtained by minimizing the mean-square distance.

Following Brockwell and Davis (1991, p.255) the likelihood function of $(Y_1,..,Y_n)$ is given by
\begin{equation}\label{Likeli}
L_n(Y_1,..,Y_n) =  (2\pi)^{-n/2} (v_0 \cdots v_{n-1})^{-1/2} \exp(-\frac{1}{2} \sum_{j=1}^n (Y_j - \hat{Y}_j)^2/v_{j-1}) .
\end{equation}
The quantity $v_j = E( Y_{j+1} - \hat{Y}_{j+1} )^2$ denotes the mean-square error. It can be recursively calculated as well (Brockwell and Davis (1991, p.172)).

Next we introduce a new CUSUM chart for the variance. It is based on the idea to apply the likelihood ratio procedure to a Gaussian process. First, we fix $n$ and consider the testing problem $H_0: \tau > n$ against $H_1: 1 \le \tau \le n$. Assuming $\Delta$ to be known the likelihood function under the null hypothesis that no change has occurred up to time $n$ is equal to the joint density of $(X_1,..,X_n)$ in the in-control case and it is given by
\begin{equation}\label{dichtein}
f_0(X_1,..,X_n) =  (2\pi)^{-n/2} (v_0 \cdots v_{n-1})^{-1/2} \exp(-\frac{1}{2} \sum_{j=1}^n (X_j - \hat{X}_j)^2/v_{j-1})
\end{equation}
with $v_j$ as above and $\hat{X}_t = \sum_{j=1}^{t-1} a_{tj} X_j$.

If there is a change at position $\tau \in \{1,...,n\}$ then the joint density of $(X_1,..,X_n)$ is equal to
\begin{align*}
f_\tau(X_1, \dotsc, X_n) & = f_0(X_1, \dotsc, X_{\tau - 1}, \frac{X_{\tau}}{\Delta}, \dotsc, \frac{X_n}{\Delta}) \times \frac{1}{\Delta^{n-\tau+1}} \\
						 & = (2\pi)^{-n/2}(v_0 \cdot \dotsm \cdot v_{n-1})^{-1/2} \frac{1}{\Delta^{n-\tau+1}}						 \exp\lbrace -\frac{1}{2} (\sum_{j=1}^{n} \frac{(Z_{j,\tau} - \widehat{Z}_{j,\tau} )^2}{v_{j-1}})\rbrace
\end{align*}
with
\[ Z_{j,\tau} = \left\{ \begin{array}{ccc} X_j & \mbox{for} & 1 \le j < \tau\\ X_j/\Delta & \mbox{for} & \tau \le j \le n \end{array} \right. , \hat{Z}_{j,\tau} = \sum_{v=1}^{j-1} a_{jv} Z_{v,\tau} . \]
Let $I_A(j)$ denote the indicator function of the set $A$ and let $T_{j,\tau} = \sum_{v=\tau}^{j-1} a_{jv} X_v$. Note that $T_{j,\tau} = 0$ for $j \le \tau$. Then
\[  \hat{Z}_{j,\tau} = \sum_{v=1}^{\min\{j-1,\tau-1\}} a_{jv} X_v + \frac{1}{\Delta} \sum_{v=\tau}^{j-1} a_{jv} X_v = \hat{X}_{j} + (\frac{1}{\Delta} - 1)  T_{j,\tau} \]
and
\[ Z_{j,\tau} - \hat{Z}_{j,\tau} = \left\{ \begin{array}{ccc} X_j - \hat{X}_j & \mbox{for} & 1 \le j < \tau\\
X_j - \hat{X}_j + (\frac{1}{\Delta} - 1) (X_j - T_{j,\tau})  & \mbox{for} & \tau \le j \le n
\end{array} \right. . \]
Thus it follows that
\begin{eqnarray}\label{dichteout}
f_\tau(X_1, \dotsc, X_n) & = & (2\pi)^{-n/2}(v_0 \cdot \cdot \cdot v_{n-1})^{-1/2} \frac{1}{\Delta^{n-\tau+1}}
\exp\left\{ -\frac{1}{2} \left(\sum_{j=1}^{\tau-1} \frac{ (X_j - \widehat{X}_{j} )^2}{v_{j-1}} \right. \right.\nonumber\\
& & + \left. \left. \sum_{j=\tau}^{n} \frac{( X_j - \hat{X}_j + (\frac{1}{\Delta} - 1) (X_j - T_{j,\tau}) )^2 }{v_{j-1}} \right) \right\} .
\end{eqnarray}
The likelihood ratio is given by
\begin{eqnarray*}
\frac{ f_0(X_1,...,X_n) }{ \max\limits_{0 \le \tau \le n} f_\tau(X_1,...,X_n) } & = & \min\{1, \min\limits_{1 \le \tau \le n} \Delta^{n-\tau+1} \\
& &  \times \exp\lbrace -\frac{1}{2} \left( \sum_{j=\tau}^{n} \frac{(X_j - \hat{X}_j)^2  - ( X_j - \hat{X}_j + (\frac{1}{\Delta} - 1) (X_j - T_{j,\tau}) )^2 }{v_{j-1}}
 \right)\rbrace \} . \end{eqnarray*}
Hence,
\begin{eqnarray*}
-2 \log\left( \frac{ f_0(X_1,...,X_n) }{ \max\limits_{0 \le \tau \le n} f_\tau(X_1,...,X_n) } \right) & = & \max\{0, \max\limits_{1 \le \tau \le n} \Bigl( - (n-\tau+1) \log(\Delta^{2}) \\
& & \hspace*{-2cm} + \left. \sum_{j=\tau}^n \frac{1}{v_{j-1}} \left[ 2 (1 - \frac{1}{\Delta}) (X_j - \hat{X}_j)(X_j-T_{j,\tau}) - (1 - \frac{1}{\Delta} )^2 (X_j - T_{j,\tau})^2  \right] \right) \} .
\end{eqnarray*}

Replacing $\Delta$ by $\Delta^* > 1$ the run length of the corresponding control chart is given by
\begin{eqnarray*}
N_{LR}(c; \Delta^*) = \inf\{ n \in I\!\!N: \max\{0, \max\limits_{1 \le i \le n} \Bigl( - (n-i+1) \log(\Delta^{*2}) \\
& & \hspace*{-9cm} + \left. \sum_{j=i}^n \frac{1}{v_{j-1}} \left[ 2 (1 - \frac{1}{\Delta^*}) (X_j - \hat{X}_j)(X_j-T_{j,i}) - (1 - \frac{1}{\Delta^*} )^2 (X_j - T_{j,i})^2  \right] \right) \} > c \}
\end{eqnarray*}
with $c >0$.

In the next section the control statistic is determined for several important special cases. Note that the above result includes all causal autoregressive moving average processes with normal white noise and all causal autoregressive fractionally integrated moving average processes with normal white noise and $|d| < 0.5$.

\subsection{Examples}

In the following we will make use of the notation

\begin{equation}\label{def}
S_n(\Delta) = \sum_{j=1}^n \frac{(X_j - \hat{X}_{j})^2}{v_{j-1}} - n K(\Delta),  \quad K(\Delta) = \frac{\log(\Delta^2)}{1-1/\Delta^2}
\end{equation}
with $\hat{X}_j$ as in Section 3.2. Note that in the case of an independent random sequence $S_n(\Delta)$ is equal to $\tilde{S}_n(\Delta)$ (see (\ref{Summe})).

The most popular family of time-correlated processes are autoregressive moving average processes (ARMA). A stochastic process is called an ARMA process of order $(p,q)$ if it is a solution of the stochastic difference equation $$Y_t =  \sum_{i=1}^p \phi_i Y_{t-i} + \varepsilon_t + \sum_{j=1}^q \theta_j \varepsilon_{t-j} .$$.

Here it is assumed that $\{\varepsilon_t\}$ is an independent and normally distributed random process with $E(\varepsilon_t)=0$ and $Var(\varepsilon_t)=\sigma^2$. Moreover, it is demanded that the roots of $1-\sum_{i=1}^p \phi_i z^i$ are all lying outside the unit circle. Then the ARMA process has a unique stationary and causal solution.

a) For an AR(1) process with $|\phi_1| < 1$ we get that $\hat{Y}_t = \phi_1 Y_{t-1}$ for $t \ge 2$ and $\hat{Y}_1 = 0$. Consequently we get that $v_t = \sigma^2$ for $t\ge 1$ and $v_0 = \gamma_0 = \sigma^2/(1-\phi_1^2)$. Furthermore $a_{t,t-1} = \phi_1$ for $t \ge 2$, $a_{tj}=0$ for $j=1,..,t-2$ and $T_{j,\tau} = \phi_1 X_{j-1} = \hat{X}_j$ for $j \ge \tau+1$ and $T_{j,\tau} = 0$ for $1 \le j \le \tau$.

We get
{\small
\[
\max\limits_{1 \le \tau \le n} \left( - (n-\tau+1) \log(\Delta^{2}) + \sum_{j=\tau}^n \frac{1}{v_{j-1}} \left[ 2 (1 - \frac{1}{\Delta}) (X_j - \hat{X}_j)(X_j-T_{j,\tau}) - (1 - \frac{1}{\Delta} )^2 (X_j - T_{j,\tau})^2  \right] \right)
\]
\[
= \max\limits_{1 \le \tau \le n} \left(- (n-\tau+1) \log(\Delta^2) + (1-\frac{1}{\Delta^2}) \sum_{j=\tau+1}^n \frac{(X_j - \hat{X}_j)^2}{v_{j-1}} + 2 (1-\frac{1}{\Delta}) \frac{X_\tau - \hat{X}_\tau}{v_{\tau-1}} X_\tau - (1-\frac{1}{\Delta})^2 \frac{X_\tau^2}{v_{\tau-1}} \right)
\]}
\begin{eqnarray}
& = & (1-\frac{1}{\Delta^2}) \max\limits_{1 \le \tau \le n} \left( S_n(\Delta) - S_\tau(\Delta) + 2 \, \frac{1}{1+1/\Delta} \, \frac{X_\tau - \hat{X}_\tau}{v_{\tau-1}} \, X_\tau - \frac{1-1/\Delta}{1+1/\Delta} \, \frac{X_\tau^2}{v_{\tau-1}} - K(\Delta) \right)\nonumber \\
%
& = & (1-\frac{1}{\Delta^2}) \max\limits_{1 \le \tau \le n} \left( S_n(\Delta) - S_\tau(\Delta) + \frac{X_\tau^2}{v_{\tau-1}} - K(\Delta) - \, \frac{2}{1+1/\Delta} \, \frac{X_\tau \hat{X}_\tau}{v_{\tau-1}}  \right) \nonumber \\
& = & (1-\frac{1}{\Delta^2}) \max\limits_{1 \le \tau \le n} \left( S_n(\Delta) - S_{\tau-1}(\Delta) - \frac{\hat{X}_\tau^2}{v_{\tau-1}}  + \, \frac{2/\Delta}{1+1/\Delta} \, \frac{X_\tau \hat{X}_\tau}{v_{\tau-1}}  \right) . \label{iid}
\end{eqnarray}
One of the problems in calculating the above quantity consists in the fact that the maximum has to be taken over all time points. This is usually quite time consuming and makes a procedures inattractive. In the present case, however, it is possible to derive a recursion. Let
\begin{equation*}
A_n^+(\Delta) = \max\limits_{1 \le \tau \le n} \left( S_n(\Delta) - S_{\tau-1}(\Delta) - \frac{\hat{X}_\tau^2}{v_{\tau-1}}  + \, \frac{2/\Delta}{1+1/\Delta} \, \frac{X_\tau \hat{X}_\tau}{v_{\tau-1}}  \right)
\end{equation*}
for $n \ge 1$ and $A_0^+(\Delta) = 0$ then it holds for $n \ge 1$ that
\begin{eqnarray*}
A_n^+(\Delta) & = & \max \left\{ \frac{(X_n - \hat{X}_n)^2}{v_{n-1}} - K(\Delta) - \frac{\hat{X}_n^2}{v_{n-1}}  + \, \frac{2/\Delta}{1+1/\Delta} \, \frac{X_n \hat{X}_n}{v_{n-1}} , \right. \\
& & \left. \max\limits_{1 \le \tau \le n-1} \left( S_n(\Delta) - S_{\tau-1}(\Delta) - \frac{\hat{X}_\tau^2}{v_{\tau-1}}  + \, \frac{2/\Delta}{1+1/\Delta} \, \frac{X_\tau \hat{X}_\tau}{v_{\tau-1}}  \right) \right\}\\
& = & \frac{(X_n - \hat{X}_n)^2}{v_{n-1}} - K(\Delta)  + \max \left\{ - \frac{\hat{X}_n^2}{v_{n-1}}  + \, \frac{2/\Delta}{1+1/\Delta} \, \frac{X_n \hat{X}_n}{v_{n-1}} ,  A_{n-1}^+(\Delta) \right\} .
\end{eqnarray*}
This representation turns out to be quite useful since it shows that the decision rule of the control scheme can be calculated recursively. Replacing $\Delta$ by $\Delta^*$ the run length of this control scheme is given by
\begin{equation}\label{NLR}
N_{LR}(c; \Delta^*) = \inf\left\{n \in \mathbb{N} :  \max\{ 0, \; A_n^+(\Delta^*)  \} > c \right\} .
\end{equation}
Putting $\phi_1 = 0$ in (\ref{iid}) we get the CUSUM variance chart for independent variables which was discussed in Section 3.2. Moreover, the representation (\ref{iid}) shows as well the relationship to residual charts. The control statistic of the CUSUM variance chart for independent samples applied to the residuals has a similar structure (cf. Section 4). The difference lies in the additional quantities based on $\hat{X}_\tau^2$ and $X_\tau \hat{X}_\tau$.\\

b) Let $\{ Y_t \}$ be a causal AR(2) process. Then $\hat{Y}_t = \phi_1 Y_{t-1} + \phi_2 Y_{t-2}$ for $t \ge 3$,
\[ \hat{Y}_2 = \frac{\phi_1}{1-\phi_2} Y_1, \quad \mbox{and} \quad \hat{Y}_1 = 0 .\]
Thus we have for $t \ge 3$ that $a_{t,t-1} = \phi_1$, $a_{t,t-2} = \phi_2$ and $a_{t,j} = 0$ for $1 \le j \le t-3$. Moreover, $a_{21} = \phi_1/(1- \phi_2)$. Furthermore, $v_0 = \gamma_0$, $v_1 = \gamma_0 (1- \phi_1^2/(1-\phi_2)^2)$, $v_t = \sigma^2$ for $t \ge 2$, and
\[ \gamma_0 = \sigma^2 \frac{1-\phi_2}{(1+\phi_2)[(1-\phi_2)^2 - \phi_1^2]} . \]
This leads to
\[ T_{j,\tau} = \left\{ \begin{array}{ccc}
\hat{X}_j & \mbox{for} & j \ge \tau+2\\
\phi_1 X_\tau & \mbox{for} & j = \tau+1 \ge 3\\
\hat{X}_2 & \mbox{for} & j=\tau + 1 = 2\\
0 & \mbox{for} & 1 \le j \le \tau
\end{array} \right. . \]
Consequently,
\begin{eqnarray*}
 \sum_{j=\tau}^n \frac{1}{v_{j-1}} \left[ 2 (1 - \frac{1}{\Delta}) (X_j - \hat{X}_j)(X_j-T_{j,\tau}) - (1 - \frac{1}{\Delta} )^2 (X_j - T_{j,\tau})^2  \right] =  \\
= (1- \frac{1}{\Delta^2}) \left( \sum_{j=\tau+1}^n \frac{(X_j - \hat{X}_j)^2}{v_{j-1}} + \frac{1}{v_{\tau-1}}(X_\tau^2-\frac{2\Delta}{1+\Delta} X_\tau \hat{X}_\tau) +  \right. \\
\left.  I_{\{2,3,..,n\}}(\tau) \; \frac{1}{v_\tau} ( \frac{2}{\Delta+1}\phi_2 X_{\tau-1}(X_{\tau+1}-\hat{X}_{\tau+1}) - \frac{\Delta-1}{\Delta+1}\phi_2^2 X_{\tau-1}^2 ) \right) \\
\end{eqnarray*}
and
{ \small
$$ \max\limits_{1 \le \tau \le n} \left( - (n-\tau+1) \log(\Delta^{2}) + \sum_{j=\tau}^n \frac{1}{v_{j-1}} \left[ 2 (1 - \frac{1}{\Delta}) (X_j - \hat{X}_j)(X_j-T_{j,\tau}) - (1 - \frac{1}{\Delta} )^2 (X_j - T_{j,\tau})^2  \right] \right) = $$
}
\begin{eqnarray*}
=(1-\frac1{\Delta^2}) \max\limits_{1 \le \tau \le n} \left( S_n(\Delta) - S_\tau(\Delta) -K(\Delta) + \frac{1}{v_{\tau-1}}(X_\tau^2-\frac{2\Delta}{1+\Delta} X_\tau \hat{X}_\tau) + \right.\\
\left. + I_{\{2,3,..,n\}}(\tau) \; \frac{1}{v_\tau} ( \frac{2}{\Delta+1}\phi_2 X_{\tau-1}(X_{\tau+1}-\hat{X}_{\tau+1}) - \frac{\Delta-1}{\Delta+1}\phi_2^2 X_{\tau-1}^2 ) \right) . \\
\end{eqnarray*}

As in the previous example, it is possible to calculate this quantity recursively. Let
\begin{eqnarray*}
B_n^+(\Delta) =\max\limits_{1 \le \tau \le n} \left( S_n(\Delta) - S_\tau(\Delta) -K(\Delta) + \frac{1}{v_{\tau-1}}(X_\tau^2-\frac{2\Delta}{1+\Delta} X_\tau \hat{X}_\tau) + \right.\\
\left. + I_{\{2,3,..,n\}}(\tau) \; \frac{1}{v_\tau} ( \frac{2}{\Delta+1}\phi_2 X_{\tau-1}(X_{\tau+1}-\hat{X}_{\tau+1}) - \frac{\Delta-1}{\Delta+1}\phi_2^2 X_{\tau-1}^2 ) \right)
\end{eqnarray*}
then
\begin{eqnarray*}
B_n^+(\Delta) = \frac{(X_n-\hat{X}_n)^2}{v_{n-1}} - K(\Delta) + \max \left( \frac{1}{v_{n-2}}(X_{n-1}^2-\frac{2\Delta}{1+\Delta} X_{n-1} \hat{X}_{n-1}) + \right.\\
\left. + I_{\{2,3,..\}}(n-1) \; \frac{1}{v_{n-1}} ( \frac{2}{\Delta+1}\phi_2 X_{n-2}(X_{n}-\hat{X}_{n}) - \frac{\Delta-1}{\Delta+1}\phi_2^2 X_{n-2}^2 ) - K(\Delta), \ B_{n-1}^+(\Delta) \right) . \\
\end{eqnarray*}

\section{A Variance Chart based on the Sequential Probability Ratio Test}
\subsection{SPRT applied to Independent Variables}

CUSUM control charts are connected to the sequential probability ratio test (SPRT). Here, we derive the CUSUM procedure directly from the related SPRT.
Assume that the variables $\{X_t\}$ are independent and identically normally distributed with expectation $\mu$.
First, we consider the simple testing problem $H_{0}^* : Var(X_t) = \gamma_0$ against $H_{1}^*: Var(X_t) = \Delta^{* 2} \gamma_0$ with known $\Delta^*$.
The SPRT says that sampling is stopped at time $n$ if $\tilde{S}_n(\Delta^*) \notin [A,B]$ with $\tilde{S}_n(\Delta)$ as in (\ref{Summe}). If $\tilde{S}_n(\Delta^*) > B$ then $H_0$ is rejected. Otherwise, if $\tilde{S}_n(\Delta^*) < A$, then $H_0$ is accepted.

Because we are interested to detect an increase in the variance we put $A=0$. Moreover, if $\tilde{S}_n(\Delta^*) \le 0$, then the chart is restarted at point zero and the procedure continues. Setting $B = c$ we get the standard CUSUM chart of Section 3.1 with run length
$$ N_{SPRT,iid}(c; \Delta^*) = \inf\{n \in \mathbb{N} : \max\limits_{0 \le i \le n} (\tilde{S}_n(\Delta^*) - \tilde{S}_i(\Delta^*) ) > c \}.$$
The decision rule can be recursively calculated by using $S_n^+(\Delta^*)$ from (\ref{Splus}). Note that it holds that $N_{SPRT,iid}(c; \Delta^*) = N_{LR,iid}(c; \Delta^*)$, i.e. the
likelihood ratio approach and the sequential probability ratio procedure lead to the same control scheme if the underlying process consists of independent random variables.

\subsection{SPRT applied to Gaussian Processes}

In this section we apply the SPRT to a Gaussian process as in Section 3.2. We consider the simple testing problem
$H_{0}^*$ against $H_{1}^*$ (see Section 4.1) with known $\Delta^*$. Using (\ref{dichtein}), (\ref{dichteout}) and the fact that $T_{j,\tau=1} = \hat{X}_j$ it follows that

\begin{eqnarray*}
\log\left( \frac{f_{\tau = 1}(X_1,X_2,\dotsc,X_n)}{f_0(X_1,X_2,\dotsc,X_n)} \right) & = & - \frac{n}{2} \log(\Delta^{* 2}) + \frac{1}{2} \; (1 - \frac{1}{\Delta^{* 2}} ) \;  \sum_{j=1}^{n} \frac{ (X_j - \hat{X}_{j} )^2}{v_{j-1}} \\
& = & \frac{1}{2} \; (1 - \frac{1}{\Delta^{* 2}} ) \; S_n(\Delta^*)
\end{eqnarray*}
with $S_n(\Delta)$ as in (\ref{def}).

Following the procedure described in the first part the SPRT leads to a CUSUM chart with run length
\begin{equation}\label{NSPRT}
N_{SPRT}(c; \Delta^*) = \inf\{n \in \mathbb{N} : \max\limits_{0 \le i \le n} (S_n(\Delta^*) - S_i(\Delta^*)) > c \}.
\end{equation}
Note that the decision rule can be calculated recursively as described in (\ref{Rekursion}). This is a great advantage of this scheme in comparison with the CUSUM schemes obtained by the likelihood approach.

Applying the SPRT approach we get a CUSUM scheme for Gaussian processes which is equal to the classical CUSUM chart for independent samples obtained in Sections 3.1 and 4.1 if the recursion is applied to the normalized residuals. Thus the chart is equal to the CUSUM residual chart for the variance.

Because the normalized residuals are independent the recursive presentation shows that the control statistic follows a Markov process. Thus for the calculation of the ARL and the average delay the Markov chain approach of Brook and Evans (1972) can be applied. Another advantage of this scheme is based on the fact that the residuals do not depend on the process parameters in the in-control state and thus the control limit does not depend on the process parameters. Consequently this approach has some advantages which simplify its application.


\section{A Variance Chart based on the Shiryaev-Roberts Approach}

The Shiryaev-Roberts (SR) approach is based on papers of Shiryaev (1963) and Roberts (1966). We make use of the change point model introduced in Section 2. In Section 3 the maximum of the likelihood ratio is taken over all possible positions of the change point, i.e. $\tau \in \{1,...,n\}$. In the SR procedure the maximum is replaced by the sum over $\tau \in \{1,...,n\}$. This procedure can be interpreted as a Bayesian procedure where $\tau$ has a geometric prior distribution with parameter $p$ converging to $0$. Pollak (1985) proved for independent variables that the SR-rule to be asymptotically Bayes risk efficient as $p \rightarrow 0$. The SR approach for independent variables has been recently discussed by several authors, e.g., Moustakides et al. (2009) and Pollak and Tartakovsky (2009).

\subsection{Shiryaev-Roberts Approach for Independent Variables}

First it is assumed that variables $\{ Y_t \}$ are independent and normally distributed with mean $\mu$ and variance $\gamma_0$.  Let $f$ denote the density of a univariate normal distribution with mean $\mu$ and variance $\gamma_0$ and let $g$ be the density of a univariate normal distribution with mean $\mu$ and variance $\Delta^2 \gamma_0$. Using the notation of Section 3 and $L_j(x) = g(x)/f(x)$ the SR statistic is given by
\begin{equation}\label{ShR1}
R_n(\Delta) = \sum\limits_{i=1}^n{\frac{f_i(X_1,...,X_n)}{f_0(X_1,...,X_n)}}=\sum\limits_{i=1}^n{\prod\limits_{j=i}^n{L_j}} = (1+R_{n-1}(\Delta)) L_n
\end{equation}
for $n\ge1$ and $R_0(\Delta) = 0$. For normal variables we get that
\begin{eqnarray}\label{ShR2}
R_n(\Delta) & = & \sum\limits_{i=1}^n {\frac1{\Delta^{n-i+1}} \exp \left\{ \frac1{2{\gamma_0}}(1-\frac1{\Delta^2})\sum\limits_{j=i}^n{{(X_j-\mu )}^2} \right\}}\nonumber\\
& = & \left( 1+R_{n-1}(\Delta) \right)\frac1\Delta \exp \left\{ \frac1{2\gamma_0}(1-\frac1{\Delta^2}){{(X_n-\mu)}^2} \right\} .
\end{eqnarray}
However, in practice, we do not know the size of the change $\Delta$. As described above it is replaced by a known quantity $\Delta^* > 1$. This leads to $R_n(\Delta^*)$.

For deriving the control statistic it was assumed that the target process is independent. Following Section 3.1 we apply this statistic for time series as well. The independent variables are replaced by the time series. The run length of the chart is given by
\begin{equation}\label{NSRiid}
N_{SR,iid}(c; \Delta^*)=\inf\{n \in \mathbb{N}: R_n(\Delta^*)>c\} .
\end{equation}

\subsection{Shiryaev-Roberts Approach for Gaussian Processes}

Now we apply this approach to a Gaussian process $\{ Y_t \}$ fulfilling the assumptions of Section 3.2. Using (\ref{dichtein}) and (\ref{dichteout}) we receive that
\begin{eqnarray*}
R_n(\Delta) &=&\sum\limits_{i=1}^n \frac{f_i(X_1, \dotsc, X_n)}{f_0(X_1, \dotsc, X_n)} \\
&=&\sum\limits_{i=1}^n \frac{1}{\Delta^{n-i+1}}\exp\left\{ -\frac12\sum_{j=i}^{n} \frac{( X_j - \hat{X}_j + (\frac{1}{\Delta} - 1) (X_j - T_{j,i}) )^2 }{v_{j-1}} + \frac12\sum_{j=i}^{n}\frac{ (X_j - \hat{X}_j )^2}{v_{j-1}} \right\} \\
&=&\sum\limits_{i=1}^n \frac{1}{\Delta^{n-i+1}}\exp\left\{ \sum_{j=i}^{n}\frac1{v_{j-1}}\left((1-\frac1\Delta)(X_j-\hat{X}_j)(X_j - T_{j,i})-\frac12(1-\frac1\Delta)^2(X_j - T_{j,i})^2 \right) \right\} .
\end{eqnarray*}
Replacing $\Delta$ by $\Delta^*$ the run length of the SR chart is given by
\begin{equation}\label{NSR}
N_{SR}(c; \Delta^*) = \inf\{n \in \mathbb{N}: R_n(\Delta^*)>c\} .
\end{equation}

\subsection{Example}
Suppose that $\{ Y_t \}$ is a causal AR(1) process. Following Example a) of Section 3.3 we get that
\begin{eqnarray*}
R_n(\Delta) &=&\sum\limits_{i=1}^n \frac{1}{\Delta^{n-i+1}} \exp\left\{ \frac12 (1-\frac1{\Delta^2})\left(\sum_{j=i}^{n}\frac{(X_j-\hat{X}_j)^2}{v_{j-1}} -\frac{\hat{X}^2_i}{v_{i-1}}+\frac2{1+\Delta}\frac{X_i\hat{X}_i}{v_{i-1}}\right)\right\}.
\end{eqnarray*}
It is possible to determine the SR statistic recursively
$$ R_n(\Delta)=\left(R_{n-1}(\Delta)+\exp\left\{(1-\frac1{\Delta^2})(\frac1{1+\Delta}\frac{X_n\hat{X}_n}{v_{n-1}}-\frac{\hat{X}^2_n}{2v_{n-1}})\right\}\right)
\frac1\Delta\exp\left\{\frac12(1-\frac1{\Delta^2})\frac{(X_n-\hat{X}_n)^2}{v_{n-1}}\right\} $$
for $n \ge 1$ and $R_0(\Delta) = 0$. As above the unknown magnitude of the change $\Delta$ is replaced by a reference value $\Delta^*$. This leads to $R_n(\Delta^*)$.


\section{A Variance Chart based on the Generalized Likelihood Ratio Approach}

The disadvantage of the procedures considered in Sections 3 to 5 consists in the fact that for the derivation of the procedures the magnitude of the change has to be known. In practice, however, in many cases no information is available about the size of a possible shift. In that case other procedures must be favored. Such approaches are discussed in Sections 6 to 8. In this section we consider the generalized likelihood ratio (GLR) approach which can be considered as an extension of the likelihood ratio method since the size of the change is assumed to be unknown. GLR charts have not received much attention in SPC up to now (cf. Reynolds and Lou (2012)) and have been mostly discussed in literature on change-point analysis (e.g., Lai (2001)).

\subsection{GLR applied to Independent Variables}

Assume that $\{Y_t\}$ is an independent normally distributed random sequence with mean $\mu$ and variance $\sigma^2$. We make use of the change point model introduced in Section 2. Let $f_{\tau,\Delta}$ denote the out-of-control density of $(X_1,..,X_n)^\prime$ then we have to calculate
\begin{equation*}
\max\limits_{1 \le \tau \le n} \sup\limits_{\Delta > 1} \log(f_{\tau,\Delta}(X_1,...,X_n)) .
\end{equation*}
We get with $\tilde{T}_n = \sum_{i=1}^n (X_i - \mu)^2/\gamma_0$ that
\[ \log(f_{\tau,\Delta}(X_1,...,X_n)) = - \frac{n}{2} \log(2\pi \gamma_0) - \frac{n-\tau+1}{2} \log(\Delta^2) - \frac{1}{2} \left[ \tilde{T}_{\tau-1} + (\tilde{T}_n - \tilde{T}_{\tau-1})/\Delta^2 \right] . \]
Let $\tilde{\Delta}_{\tau,n}^2 = (\tilde{T}_n - \tilde{T}_{\tau-1})/(n-\tau+1)$ and let $\Delta_{\tau,n}^{2} = \max\{1, \tilde{\Delta}_{\tau,n}^2 \}$. It holds that
\[  \sup\limits_{\Delta > 1} \log(f_{\tau,\Delta}(X_1,...,X_n)) = \log(f_{\tau,\Delta_{\tau,n}}(X_1,...,X_n)) \]
since the derivative of $\log(f_{\tau,\Delta}(X_1,...,X_n))$ with respect to $\Delta^2$ is positive if $\Delta^2 < \tilde{\Delta}_{\tau,n}^2$ and it is negative for $\Delta^2 > \tilde{\Delta}_{\tau,n}^2$. It holds that\\

$
\log(f_{\tau,\Delta_{\tau,n}}(X_1,...,X_n))
$
\begin{equation*}
= \left\{ \begin{array}{ccc}
- \frac{n}{2} \, \log(2\pi \gamma_0) - \frac{1}{2} [\tilde{T}_{\tau-1}+n-\tau+1]- \frac{n-\tau+1}{2}  \, \log(\tilde{\Delta}_{\tau,n}^2) & \mbox{if} & \tilde{\Delta}_{\tau,n}^2 \ge 1\\
- \frac{n}{2} \log(2\pi \gamma_0) - \frac{\tilde{T}_n}{2} & \mbox{if} & \tilde{\Delta}_{\tau,n}^2 < 1
\end{array} \right. .
\end{equation*}
Thus
\begin{equation*}
\sup\limits_{\Delta>1} \log\left( \frac{f_{\tau,\Delta}(X_1,...,X_n)}{f_0(X_1,...,X_n)} \right) =
\frac{n-\tau+1}2 [ \Delta_{\tau,n}^2 -1 - \log(\Delta_{\tau,n}^2) ] .
\end{equation*}
Because $x-1-\log(x) \ge 0$ for $x \ge 1$ we obtain that
\begin{equation*}
2 \;  \max\limits_{1 \le \tau \le n+1} \sup\limits_{\Delta>1} \log\left( \frac{f_{\tau,\Delta}(X_1,...,X_n)}{f_0(X_1,...,X_n)} \right) =  \max\limits_{1 \le \tau \le n} \left\{ (n-\tau+1) \; \left[ \Delta_{\tau,n}^2 -1 - \log(\Delta_{\tau,n}^2) \right] \right\} .
\end{equation*}
Consequently the stopping rule of the GLR test is given by
\[ N_{GLR,iid}(c) = \inf\{n \in \mathbb{N} : \max\limits_{1 \le i \le n} \; \left\{ (n-i+1) \left[ \Delta_{i,n}^2 -1 - \log(\Delta_{i,n}^2) \right] \right\} > c \}. \]

\subsection{GLR applied to Gaussian Processes}

Suppose that $\{ Y_t \}$ is a Gaussian process with mean zero and covariance function $k(i,j) = Cov(X_i, X_j)$ fulfilling the conditions of Section 3.2. Using $T_n = \sum_{i=1}^n (X_i - \hat{X}_i)^2/v_{i-1}$ and (\ref{dichteout}) it follows that
\begin{eqnarray*}
\log(f_{\tau,\Delta}(X_1, \dotsc, X_n)) & = &  const - \frac{n-\tau+1}{2} \; \log(\Delta^2) \\
& &- \frac{1}{2} \left( T_{\tau-1}  + \sum_{j=\tau}^{n} \frac{( X_j - \hat{X}_j + (\frac{1}{\Delta} - 1) (X_j - T_{j,\tau}) )^2 }{v_{j-1}} \right) .
\end{eqnarray*}
Now we will maximize the function $\log \; f_{\tau,\Delta}(X_1, \dotsc, X_n)$ with respect to $\Delta$. For that reason we calculate the derivative
\begin{eqnarray*}
\frac{d}{d\Delta} \log(f_{\tau,\Delta}(X_1, \dotsc, X_n)) &=& - \frac{n-\tau+1}{\Delta} + \frac{1}{\Delta^2}\sum_{j=\tau}^{n} \frac{( X_j - \hat{X}_j + (\frac{1}{\Delta} - 1) (X_j - T_{j,\tau}) ) }{v_{j-1}}(X_j - T_{j,\tau}) \\
& = & - \frac{n-\tau+1}{\Delta} + \frac{1}{\Delta^2} (\dot{S}_{n,\tau} - \ddot{S}_{n,\tau}) + \frac{1}{\Delta^3}\ddot{S}_{n,\tau}
\end{eqnarray*}
where
\begin{equation}\label{SPunkt}
\dot{S}_{n,\tau} = \sum_{j=\tau}^{n} \frac{( X_j - \hat{X}_j) (X_j - T_{j,\tau}) }{v_{j-1}} , \ \
\ddot{S}_{n,\tau} = \sum_{j=\tau}^{n} \frac{ (X_j - T_{j,\tau})^2 }{v_{j-1}}  .
\end{equation}
Let
\[ \dot{\Delta}_{\tau,n} = \frac{ \dot{S}_{n,\tau} - \ddot{S}_{n,\tau} + \sqrt{ ( \dot{S}_{n,\tau} - \ddot{S}_{n,\tau} )^2 + 4(n-\tau+1)\ddot{S}_{n,\tau} } }{2(n-\tau+1)} . \]
Because the derivative of $ \log(f_{\tau,\Delta}(X_1,...,X_n))$ with respect to $\Delta$ is positive if $0 < \Delta < \dot{\Delta}_{\tau,n} $
and it is negative else, it holds that
\[  \sup\limits_{\Delta > 1} \log(f_{\tau,\Delta}(X_1,...,X_n)) = \log(f_{\tau,\Delta^*_{\tau,n}}(X_1,...,X_n)) \]
with $\Delta^*_{\tau,n} = \max\{1, \dot{\Delta}_{\tau,n} \}$.
Thus
\begin{eqnarray}\label{GLR}
\sup\limits_{\Delta > 1} \log\left( \frac{ f_{\tau,\Delta}(X_1,...,X_n)}{f_0(X_1,...,X_n)} \right) & &\nonumber \\
& \hspace*{-6cm} = & \hspace*{-3cm} -(n-\tau+1) \log(\Delta^*_{\tau,n}) - \frac12(\frac{1}{\Delta^*_{\tau,n}}-1)(2\dot{S}_{n,\tau} + (\frac{1}{\Delta^*_{\tau,n}}-1) \ddot{S}_{n,\tau}) .
\end{eqnarray}

Consequently the stopping rule for this chart is given by the following
$$N_{GLR}(c) = \inf \bigg\{ n \in \mathbb{N} : \max\limits_{1 \le i \le n} \Big\{ -(n-i+1) \log(\Delta^*_{i,n}) $$
$$ - \frac12(\frac{1}{\Delta^*_{i,n}}-1)(2\dot{S}_{n,i} + (\frac{1}{\Delta^*_{i,n}}-1) \ddot{S}_{n,i}) \Big\} > c \bigg\} .$$

\subsection{Example}

Let $\{Y_t\}$ be a causal AR(1) process. Using the results of Section 3.3a) we get that
\begin{eqnarray}
\dot{S}_{n,\tau} & = & \sum_{j=\tau}^{n} \frac{( X_j - \hat{X}_j) (X_j - T_{j,\tau}) }{v_{j-1}} =
T_n - T_\tau + \frac{(X_\tau - \hat{X}_\tau) X_\tau}{v_{\tau-1}} \label{Summe_1} , \\
\ddot{S}_{n,\tau} & = & \sum_{j=\tau}^{n} \frac{ (X_j - T_{j,\tau})^2 }{v_{j-1}} = T_n - T_\tau + \frac{X_\tau^2}{v_{\tau-1}} \label{Summe_2} .
\end{eqnarray}
Thus $ \dot{S}_{n,\tau} = \ddot{S}_{n,\tau} -\frac{X_\tau \hat{X}_\tau}{v_{\tau-1}} . $
Consequently
\begin{eqnarray*}
\dot{\Delta}_{\tau,n} & = & \frac{ \dot{S}_{n,\tau} - \ddot{S}_{n,\tau} + \sqrt{ ( \dot{S}_{n,\tau} - \ddot{S}_{n,\tau} )^2 + 4(n-\tau+1)\ddot{S}_{n,\tau} } }{2(n-\tau+1)} \\
& = & \frac{ -\frac{X_\tau \hat{X}_\tau}{v_{\tau-1}} + \sqrt{ ( \frac{X_\tau \hat{X}_\tau}{v_{\tau-1}} )^2 + 4(n-\tau+1)( T_n - T_\tau +\frac{X^2_\tau}{v_{\tau-1}} ) } }{2(n-\tau+1)}
\end{eqnarray*}
and the stopping rule is given by
$$N_{GLR}(c)=\inf \biggl\{ n \in \mathbb{N}: \max\limits_{1 \le i \le n} \Bigl\{-(n-i+1) \log(\Delta^*_{i,n}) - \frac12(\frac1{\Delta^*_{i,n}}-1)  $$
$$ \biggl((\frac1{\Delta^*_{i,n}}+1)(T_n - T_i+\frac{X^2_i}{v_{i-1}})-2\frac{X_i \hat{X}_i}{v_{i-1}}\biggl) \Bigl\} > c \biggl\} .$$



\section{A Variance Chart based on the Generalized SPRT}

Let $\{ Y_t \}$ be a Gaussian process as assumed in Section 3.2. Following the SPRT approach sketched in Section 4 and using (\ref{GLR}) we get
\begin{eqnarray*}
\sup\limits_{\Delta > 1} \log\left( \frac{ f_{\tau=1,\Delta}(X_1,...,X_n)}{f_0(X_1,...,X_n)} \right) & &\nonumber \\
& \hspace*{-6cm} = & \hspace*{-3cm} -n \log(\Delta^*_{1,n}) - \frac12(\frac{1}{\Delta^*_{1,n}}-1)(2\dot{S}_{n,1} + (\frac{1}{\Delta^*_{1,n}}-1) \ddot{S}_{n,1}) .
\end{eqnarray*}
Since $\dot{S}_{n,1} = \ddot{S}_{n,1} = T_n$ we get that $\dot{\Delta}_{1,n} = \sqrt{T_n/n}$ and thus
\begin{eqnarray*}
\sup\limits_{\Delta > 1} \log\left( \frac{ f_{\tau=1,\Delta}(X_1,...,X_n)}{f_0(X_1,...,X_n)} \right) & &\nonumber \\
& \hspace*{-6cm} = & \hspace*{-3cm} \left\{ \begin{array}{ccc}
-n \left[ \log(T_n/n) - T_n/n + 1\right] /2 & \mbox{if} & T_n/n \ge 1\\
0 & \mbox{if} & T_n/n < 1
\end{array} \right. \; = \; h_n(T_n/n)
\end{eqnarray*}
with
\begin{equation}\label{hn}
h_n(x) = n h(x) = n (x-1-log(x))/2 .
\end{equation}
Note that $h_n( T_n/n ) \ge 0$. Following Section 4 a control chart is obtained by applying this approach sequentially. The run length of this scheme is
$$ N_{GSPRT}(c) =  inf\{ n \in I\!\!N:  \max\limits_{0 \le i \le n}( h_n( T_n/n) ) - h_i( T_i/i ) ) > c \} . $$
Assuming $\tau = 1$ in (\ref{model}) it holds that $P_{\tau=1,\Delta}( T_n \le x ) = \chi_n^2(x/\Delta^2)$ and thus $P_{\tau=1,\Delta}( T_n/n < 1 )
= \chi_n^2(n/\Delta^2) \rightarrow 0$ if $\Delta \rightarrow \infty$. This means that the probability that $h_n(T_n/n)$ is positive is increasing with $\Delta$.

\section{Variance Charts based on the Generalized Shiryaev-Roberts Approach}

\subsection{GSR applied to Independent Variables}

Assume that the variables $\{ Y_t \}$ are independent and normally distributed with mean $\mu$ and variance $\gamma_0$. Following (\ref{ShR1}) and
(\ref{ShR2}) it is necessary to determine the maximize of $R_n(\Delta)$ over $\Delta$. However, it is not possible to get an explicit expression for the value of $\Delta$ which maximizes this quantity since the derivation of $R_n(\Delta)$ with respect to $\Delta$ leads to an exponential sum which is difficult to handle. For that reason we choose another procedure. Instead of $R_n(\Delta)$ we consider
$$ R_{n}^*(\Delta) = \sum\limits_{i=1}^n \log\left( \frac{f_{i,\Delta}(X_1,...,X_n)}{f_0(X_1,...,X_n)} \right) . $$
Because the logarithm is a strictly increasing continuous function this means that instead of the arithmetic mean the geometric mean is maximized.

In the case of independent variables we get that
\begin{eqnarray*}
R^*_{n,iid}(\Delta) & = & \log\left( \prod\limits_{i=1}^n \frac{f_{i,\Delta}(X_1,...,X_n)}{f_0(X_1,...,X_n)} \right)\\
& = & \sum_{i=1}^n \left( -(n-i+1) \log(\Delta) + (1 - \frac{1}{\Delta^2}) (\tilde{T}_n - \tilde{T}_{i-1}) \right) .
\end{eqnarray*}
Determining the derivative of $R_{n,iid}^*(\Delta)$ we see that the maximum of $R_{n,iid}^*(\Delta)$ is attained at
$$ \tilde{\Delta}_{n,iid}^2 = \frac{2}{n(n+1)} U_n $$
with
$$ U_n = \sum\limits_{i=1}^n (\tilde{T}_n - \tilde{T}_{i-1}) = \sum_{i=1}^n i \frac{(X_i - \mu)^2}{\gamma_0}. $$
Because $R_{n,iid}^*(\Delta)$ is a concave function the maximum over $\Delta \ge 1$ is attained at $\hat{\Delta}_{n,iid} = \max \{1,\tilde{\Delta}_{n,iid} \}$.
For $\tilde{\Delta}_{n,iid} \le 1$ it follows that $\sup\limits_{\Delta \ge 1} R_{n,iid}^*(\Delta) = 0$, else
\begin{eqnarray*}
 2 \, \sup\limits_{\Delta \geq 1} R_{n,iid}^*(\Delta) & = &-\frac{n(n+1)}2 \log(\tilde{\Delta}_{n,iid}^2) + (1-1/\tilde{\Delta}_{n,iid}^2) U_n \\
 &=& U_n - \frac{n(n+1)}2-\frac{n(n+1)}2 \left( \log( U_n )  - \log\left( \frac{n(n+1)}2 \right) \right) \\
 & = & \frac{n(n+1)}{2} \left( \tilde{\Delta}_{n,iid}^2 - 1 - log( \tilde{\Delta}_{n,iid}^2 ) \right) \\
 & = & h_{n(n+1)}( \tilde{\Delta}_{n,iid}^2 )
\end{eqnarray*}
with $h_n$ as in (\ref{hn}). The run length of the control chart is given by
$$N_{GSR,iid}(c) = \inf \{n \in \mathbb{N}: h_{n(n+1)}(\hat{\Delta}_{n,iid}^2) > c \} . $$

\subsection{GSR for Gaussian Processes}

Now let $\{ Y_t \}$ be a Gaussian process as assumed in Section 3.2. As in the previous section we make use of the geometric mean $R^*_n(\Delta)$. In the present case we get that
\begin{eqnarray*}
2 \, R^*_n(\Delta) & = & \sum\limits_{k=1}^n \left( -(n-k+1)\log(\Delta^2) + 2(1-\frac1\Delta)\dot{S}_{n,k} - (1-\frac1\Delta)^2\ddot{S}_{n,k}\right)\\
& = & - \frac{n(n+1)}{2} \, log(\Delta^2) + 2 (1-\frac1\Delta) \dot{U}_n - (1-\frac1\Delta)^2 \ddot{U}_n
\end{eqnarray*}
with
$$ \dot{U}_n = \sum_{k=1}^n \dot{S}_{n,k}, \quad \ddot{U}_n = \sum_{k=1}^n \ddot{S}_{n,k} $$
and $\dot{S}_{n,k}$ and $\ddot{S}_{n,k}$ as in (\ref{SPunkt}). $R^*_n(\Delta)$ is a concave function for $\Delta > 0$ and its maximum for $\Delta > 0$ is attained at position
$$ \tilde{\Delta}_n = \frac{\dot{U}_n - \ddot{U}_n + \sqrt{ (\dot{U}_n - \ddot{U}_n)^2 + 2n(n+1) \ddot{U}_n } }{n(n+1)} .$$
Thus it holds that
$$ 2 \, \sup\limits_{\Delta \ge 1} R^*_n(\Delta) = -\frac{n(n+1)}2\log(\hat{\Delta}_n^2) + 2(1-\frac1{\hat{\Delta}_n}) \dot{U}_n - \frac12(1-\frac{1}{\hat{\Delta}_n})^2 \ddot{U}_n = g_n(\dot{U}_n, \ddot{U}_n)$$
where $\hat{\Delta}_n = \max\{1, \tilde{\Delta}_n \}$. The run length of this scheme is
$$ N_{GSR}(c) = inf\{ n \in I\!\!N: g_n(\dot{U}_n, \ddot{U}_n) > c \} . $$

\section{Comparison Study}

In the above sections several new control schemes for detecting a change in the variance of a time series were introduced. Because no optimality results are known our aim is to give a practitioner a clear recommendation which chart should be applied in a specific situation. In order to compare control charts at all, we have to calibrate them. The control limits of all charts are chosen such that their in-control ARLs are the same. Here we fix the in-control ARL to be $500$. Using the corresponding control limits the out-of-control ARLs of all charts are compared with each other using the out-of-control ARL $E_{\tau=1, \Delta}(N(c))$ as well as the average delay $E_{\tau, \Delta}(N(c)-\tau+1 | N(c) \ge \tau)$. These performance criteria are mostly applied in literature.

In the present case we do not have explicit formulas for the calculation of the performance criteria. Such formulas are not available for control charts for time-dependent processes. Here we make use of an extensive simulation study. In each case the average run length and the average delay were determined within a simulation study based on $10^6$ repetitions. The only exception are the GLR charts where no recursive presentation was given and the calculation of the average run length and the average delay is more time consuming. In that case we used $10^5$ repetitions.

Note that some control charts depend on a reference value $\Delta^* > 1$.  In our study $\Delta^*$ is taking values within the set $\{1.10, 1.20, 1.30, 1.40, 1.50, 1.75, 2.00, 2.25, 2.50, 2.75, 3.00\}$.

In our comparison study the target process is an AR(1) process with standard normally distributed white noise. The coefficients of the process are taking values within
$\{-0.9, -0.8,-0.7,\dotsc,0.7,0.8,0.9\}$.

In Tables 1 to 3 the out-of-control ARLs of the considered charts are given. Because the ARL turns out to be symmetric with respect to the coefficient $\phi_1$ of the AR process we only show the results for non-negative values of $\phi_1$. In each row and each column the ARLs of nine control charts are given, above the variance chart for iid variables (Section 3.1, cf. (\ref{Niid})), followed by the LR chart (Section 3.2, cf. (\ref{NLR})), the SPRT chart (Section 4.2, cf. (\ref{NSPRT})), the Shiryaev-Roberts chart for iid variables applied to time series (Section 5.1, cf. (\ref{NSRiid})), the Shiryaev-Roberts chart for Gaussian processes (Section 5.2, cf. (\ref{NSR})), the GLR chart of Section 6.2, the GSPRT chart of Section 7, and the GSR chart of Section 8.2. The first five charts depend on a reference value. For these charts the smallest out-of-control ARL over all $\Delta^*$ is listed. In parenthesis the value of $\Delta^*$ is given where the minimum is attained. The other four charts are generalized schemes and do not depend on a reference value. In Tables 1 to 3 the ARLs of all charts are written in bold which for a fixed value of $\Delta$ deviate from the smallest out-of-control ARL by only $2 \%$.

\begin{center}
[ Tables 1 to 3 about here. ]
\end{center}

The results of the comparison study are very interesting. First, it can be seen that the results for the charts based on the independence assumption, i.e. the schemes of Section 3.1 and 5.1, are getting worse if the correlation structure of the target process increases. Since the other schemes behave much better they should not be applied. Second, the minimum out-of-control ARL of the LR chart and the SPRT chart is always attained if $\Delta^*$ is equal to the true value of the change $\Delta$. For the SR chart a slightly different behavior is observed. Here the best value is greater or equal to $\Delta$. If the optimal values for the LR and the SPRT chart are taken they provide smaller ARLs than the SR chart. Among the five schemes with a reference value the LR and the SPRT chart behave the best. Their results are very similar.

The analysis of the generalized charts shows that again the chart based on the independence assumption (here: GSRiid, Section 8.1) behaves bad. The results for the GSPRT chart are also not very good for small changes but it is the best generalized scheme for $\Delta \ge 1.75$. The best results for small changes ($\Delta \le 1.3$) are obtained for the GSR chart. The chart even behaves better than all charts using a reference parameter. This is very remarkable. However, for $\Delta \ge 1.4$ the LR chart and the SPRT chart dominate this scheme.

In Figure 1 we discuss the sensitivity of the LR, the SPRT, and the SR chart with respect to the choice of the reference value. In practice we usually do not know the magnitude of the expected change in advance. In the figure the results of the GLR, the GSPRT, and the GSR chart are given as well. We focus on two changes, $\Delta = 1.3$ and $\Delta = 2.0$ and on two values of the coefficient of the AR(1) process, $\phi_1 = 0.4$ and $\phi_1 = 0.8$. The figure shows the dominance of the GSR chart for $\Delta = 1.3$. For a small correlation structure, here $\phi_1 = 0.4$, the GLR chart and the SPRT chart show to be better than the GLR chart if $\Delta^* \le 2.0$. This means that if the deviation from the optimal $\Delta = 1.3$ is not too large then these charts have a smaller ARL than the GLR chart but if the deviation is large ($\Delta^* \geq 2.25$) then the GLR scheme must be preferred. A similar behavior can be observed for $\phi_1 = 0.8$. However, if the change is larger, here $\Delta = 2.0$, then the LR and the SPRT chart are always
better than the GLR chart. Even if the reference value is chosen completely different than the true value of the change the schemes are better. Only the GSPRT chart turns out to be better than the LR and the SPRT scheme if the deviation from the optimal value is sufficiently large. It is interesting as well that it seems to be better to overestimate the value of the change than to underestimate it.

\begin{center}
[ Figure 1 about here. ]
\end{center}

Up to now the charts were compared using the average run length. For this performance measure it is assumed that the change already arises at the beginning, i.e. $\tau = 1$. This is of course a restriction. The average delay is a more general criteria because the change may arise at any position. In Table 4 it is assumed that the change arises up to the 50th observation, i.e. $1 \le \tau \le 50$. Moreover, we focus on the changes $\Delta = 1.3$ and $\Delta = 2.0$. The values of the LR, the SPRT, and the SR chart refer to the optimal choice of the reference value. The table shows that except for the GSPRT chart the worst average delay is attained at $\tau = 1$, i.e. it is equal to the ARL. As $\tau$ increases the average delay is decreasing and it does not change a lot for $\tau \ge 20$. The GSPRT chart is an exception. The worst average delay is attained at $\tau = 50$ and its minimum ARL is observed for a small value of $\tau$. The table illustrates that the GSR chart behaves quit well for small changes. The limit of the average delay and the worst case average delay are the smallest ones for this chart. However, for large changes its behavior is more complicate. While its ARL is the largest one it turns out to be more effective if the change arises at a later time point.

 \begin{center}
[ Table 4 about here. ]
\end{center}

\section{Summary}

In this paper several new control charts for the detection of a change in a Gaussian process are introduced. The charts are derived by making use of the likelihood ratio approach, the sequential probability ratio test, and the Shiryaev-Roberts approach. For the derivation of the charts it is assumed that the size of the change is known. The obtained charts depend on a reference value which has to be chosen suitably. We consider the case of an unknown size of the change as well. This attempt leads to generalized control charts.

In an extensive simulation study we compare the introduced control charts with each other. The target process is assumed to be an AR(1) process. Using the ARL as a performance criterion it turns out that the GSR chart behaves the best for small changes ($\Delta \le 1.3$). For detecting medium and large changes ($\Delta > 1.3$) the LR and the SPRT chart turn out to be better. They depend on an additional reference value. The minimum out-of-control ARL is obtained for both schemes if the reference value is chosen equal to the true value of the change. It turns out that for larger changes the LR and the SPRT chart are still better than the best generalized chart if the true change is not dramatically underestimated.

If we analyze the control charts using the average delay it can be seen that except the GSPRT chart the worst average delay is always attained at $\tau = 1$, i.e. it is equal to the ARL. The GSR chart provides the best results for a small change. For medium and large change the LR and the SPRT chart must be favored if the change is expected to be at the beginning of the monitoring process. However, if it appears at a later time point the GSR turns out to have the smallest delay.

\section{Acknowledgment}

The authors wish to express their gratitude to Matthias Lech (Department of Supply Chain Management, European University Viadrina) for valuable discussions, comments, and remarks.



\newpage

\begin{sidewaystable}\small
\caption{Out-of-control ARLs of several CUSUM control charts (variance chart for iid observations, LR chart, SPRT chart, Shiryaev-Roberts charts, GLR chart, GSPRT chart, and Generalized Shiryaev-Roberts charts) for an in-control ARL of $500$}
\vspace*{0.3cm}
\begin{tabular}{|c|c|c|c|c|c|c|c|c|c|c|}
\hline
$\Delta/ \phi_1$
     &   0            & 0.1          & 0.2          & 0.3         & 0.4          & 0.5          & 0.6          & 0.7          & 0.8          & 0.9         \\
     \hline
1.10 &   116.75(1.10) & 118.20(1.10) & 123.36(1.10) & 129.56(1.20)& 139.29(1.30) & 152.69(1.50) & 168.85(1.75) & 190.95(2.25) & 233.90(3.00) & 363.45(3.00)\\
     &   116.76(1.10) & 116.84(1.10) & 116.87(1.10) & 117.07(1.10)& 117.16(1.10) & 117.11(1.10) & 117.49(1.10) & 117.75(1.10) & 118.10(1.10) & 120.28(1.10)\\
     &   116.89(1.10) & 116.88(1.10) & 116.96(1.10) & 117.06(1.10)& 117.08(1.10) & 117.04(1.10) & 117.22(1.10) & 117.34(1.10) & 117.46(1.10) & 117.66(1.10)\\
     &  125.43(1.30)  & 126.25(1.30) & 130.03(1.30) & 135.90(1.30)& 143.91(1.40) & 154.39(1.50) & 169.31(1.75) & 191.05(2.25) & 233.56(3.00) & 362.92(3.00)\\
     &   125.55(1.30) & 125.34(1.30) & 125.37(1.30) & 125.48(1.30)& 125.03(1.30) & 125.67(1.30) & 125.57(1.30) & 125.53(1.30) & 125.63(1.30) & 125.97(1.30)\\
     &   121.65       & 121.66       & 121.75       & 121.83      & 121.97       & 122.13       & 122.47       & 122.54       & 122.95       & 123.76      \\
     &   351.98       & 353.76       & 353.46       & 354.41      & 355.07       & 355.05       & 358.02       & 356.63       & 358.36       & 359.05      \\
     & \textbf{77.84} &  95.68       & 139.16       & 192.72      & 248.62       & 299.70       & 342.69       & 378.18       & 406.98       & 430.80      \\
     & \textbf{79.06} &\textbf{ 79.04} & \textbf{79.08} & \textbf{79.12} & \textbf{79.16} & \textbf{79.24} & \textbf{ 79.29} & \textbf{79.44} & \textbf{79.80 } & \textbf{80.51} \\
     \hline
1.20 &   54.20(1.20)  & 54.99(1.20)  & 57.84(1.20)  & 61.90(1.30) & 68.23(1.40)  & 77.21(1.50)  & 89.38(2.00)  & 104.74(2.50) & 146.93(3.00) & 281.08(3.00)\\
     &   54.20(1.20)  & 54.09(1.20)  & 54.18(1.20)  & 54.35(1.20) & 54.40(1.20)  & 54.62(1.20)  & 54.94(1.20)  & 55.11(1.20)  & 55.54(1.20)  &  56.80(1.20)\\
     &   54.15(1.20)  & 54.15(1.20)  & 54.23(1.20)  & 54.32(1.20) & 54.41(1.20)  & 54.53(1.20)  & 54.68(1.20)  & 54.83(1.20)  & 54.95(1.20)  &  55.06(1.20)\\
     &   58.69(1.40)  & 59.34(1.40)  & 61.74(1.40)  & 65.78(1.40) & 70.91(1.50)  & 79.21(1.75)  & 89.69(2.00)  & 104.92(2.50) & 147.13(3.00) & 281.11(3.00)\\
     &   58.42(1.40)  & 58.61(1.40)  & 58.79(1.40)  & 58.58(1.40) & 58.54(1.40)  & 58.99(1.50)  & 58.77(1.40)  &  58.87(1.40) & 58.99(1.40)  &  58.97(1.40)\\
     &   61.43        & 61.44        & 61.49        & 61.59       & 61.72        & 61.89        & 61.90        &  62.38       &  62.73       &  63.51      \\
     &   99.68        & 99.49        & 99.33        & 99.43       & 99.24        & 99.69        & 100.77       & 100.34       & 101.91       & 102.92      \\
     &   \textbf{45.45}        & 55.97        & 83.18        & 121.30      & 167.18       & 215.58       & 262.36       & 305.43       & 344.07       & 379.30      \\
     &   \textbf{45.70}        & \textbf{45.69}        & \textbf{45.71}        & \textbf{45.75}       & \textbf{45.82}        & \textbf{45.88}        & \textbf{45.99}        &  \textbf{46.19}       &  \textbf{46.62}       &  \textbf{47.22}      \\
     \hline
1.30 &   \textbf{32.32}(1.30)  & \textbf{32.86}(1.30)  & 34.69(1.30)  & 37.60(1.40) & 41.78(1.50)  & 47.97(1.75)  & 56.16(2.00)  & 68.10(2.75)  & 105.21(3.00) & 226.55(3.00)\\
     &   \textbf{32.32}(1.30)  & \textbf{32.30}(1.30)  & \textbf{32.39}(1.30)  & \textbf{32.50}(1.30) & \textbf{32.66}(1.30)  & \textbf{32.82}(1.30)  & \textbf{33.02}(1.30)  & \textbf{33.24}(1.30)  & \textbf{33.70}(1.30)  &  34.58(1.30)\\
     &   \textbf{32.32}(1.30)  & \textbf{32.32}(1.30)  & \textbf{32.39}(1.30)  & \textbf{32.47}(1.30) & \textbf{32.57}(1.30)  & \textbf{32.78}(1.30)  & \textbf{32.94}(1.30)  & \textbf{33.11}(1.30)  & \textbf{33.24}(1.30)  &  \textbf{33.27}(1.30)\\
     &   35.14(1.50)  & 35.68(1.50)  & 37.24(1.50)  & 40.29(1.75) & 43.87(1.75)  & 48.94(1.75)  & 56.76(2.00)  & 68.42(2.75)  & 105.37(3.00) & 226.47(3.00)\\
     &   35.23(1.50)  & 35.06(1.50)  & 35.10(1.50)  & 35.19(1.50) & 35.27(1.50)  & 35.19(1.50)  & 35.34(1.50)  & 35.41(1.50)  & 35.37(1.50)  &  35.51(1.50)\\
     &   39.11        & 39.11        & 39.18        & 39.28       & 39.42        & 39.59        & 39.81        & 40.04        & 40.39        &  41.11      \\
     &   48.97        & 49.02        & 49.08        & 49.19       & 49.14        & 49.24        & 49.87        & 49.95        & 50.69        & 51.08       \\
     &   \textbf{32.54}        & 40.07        & 59.95        & 89.21       & 126.61       & 169.15       & 213.21       & 256.96       &298.69        & 339.33      \\
     &   \textbf{32.54}        & \textbf{32.55}        & \textbf{32.57}        & \textbf{32.61}       & \textbf{32.66}        & \textbf{32.74}        & \textbf{32.83}        & \textbf{33.01}        & \textbf{33.39}        &  \textbf{33.96}      \\
     \hline

\end{tabular}
\end{sidewaystable}

\begin{sidewaystable}\small
\caption{Out-of-control ARLs of several CUSUM control charts (variance chart for iid observations, LR chart, SPRT chart, Shiryaev-Roberts charts, GLR chart, GSPRT chart and Generalized Shiryaev-Roberts charts) for an in-control ARL of $500$}
\begin{tabular}{|c|c|c|c|c|c|c|c|c|c|c|}
\hline
$\Delta/ \phi_1$
     &   0            & 0.1          & 0.2          & 0.3         & 0.4          & 0.5          & 0.6          & 0.7          & 0.8          & 0.9         \\
     \hline
1.40 &   \textbf{22.03}(1.40)  & \textbf{22.47}(1.40)  & 23.71(1.40)  & 25.82(1.50) & 29.18(1.75)  & 33.47(1.75)  & 39.73(2.25)  & 48.84(3.00)  & 81.10(3.00)  & 188.12(3.00)\\
     &   \textbf{22.04}(1.40)  & \textbf{22.04}(1.40)  & \textbf{22.08}(1.40)  & \textbf{22.21}(1.40) & \textbf{22.35}(1.40)  & \textbf{22.51}(1.40)  & \textbf{22.72}(1.40)  & \textbf{22.99}(1.40)  & \textbf{23.27}(1.40)  &  23.94(1.40)\\
     &   \textbf{22.03}(1.40)  & \textbf{22.03}(1.40)  & \textbf{22.10}(1.40)  & \textbf{22.18}(1.40) & \textbf{22.29}(1.40)  & \textbf{22.43}(1.40)  & \textbf{22.60}(1.40)  & \textbf{22.75}(1.40)  & \textbf{22.90}(1.40)  &  \textbf{22.99}(1.40)\\
     &   23.77(1.75)  & 24.11(1.75)  & 25.33(1.75)  & 27.37(1.75) & 30.22(1.75)  & 34.33(2.00)  & 40.15(2.25)  & 49.01(3.00)  & 81.36(3.00)  & 187.98(3.00)\\
     &   23.69(1.75)  & 23.76(1.75)  & 23.88(1.75)  & 23.78(1.75) & 24.02(1.75)  & 24.02(1.75)  & 24.21(1.75)  & 24.18(1.75)  & 24.39(1.75)  & 24.48(1.75) \\
     &   27.97        & 27.98        & 28.05        & 28.14       & 28.26        & 28.43        & 28.60        & 28.89        & 29.23        & 29.84       \\
     &   30.44        & 30.48        & 30.60        & 30.60       & 30.64        & 30.76        & 31.03        & 31.04        & 31.54        & 31.88       \\
     &   25.56        & 31.37        & 47.13        & 70.91       & 102.44       & 139.67       & 180.25       & 222.23       & 264.49       & 307.27      \\
     &   25.42        & 25.43        & 25.45        & 25.49       & 25.54        & 25.62        & 25.74        & 25.90        & 26.20        & 26.72       \\
     \hline
1.50 &   \textbf{16.30}(1.50)  & \textbf{16.61}(1.50)  & 17.61(1.50)  & 19.33(1.75) & 21.65(1.75)  & 25.10(2.00)  & 30.04(2.25)  & 37.37(3.00)  & 65.60(3.00)  & 159.87(3.00)\\
     &   \textbf{16.30}(1.50)  & \textbf{16.32}(1.50)  & \textbf{16.40}(1.50)  & \textbf{16.51}(1.50) & \textbf{16.61}(1.50)  & \textbf{16.78}(1.50)  & \textbf{16.97}(1.50)  & \textbf{17.20}(1.50)  & \textbf{17.48}(1.50)  & 17.95(1.50) \\
     &   \textbf{16.31}(1.50)  & \textbf{16.33}(1.50)  & \textbf{16.37}(1.50)  & \textbf{16.45}(1.50) & \textbf{16.57}(1.50)  & \textbf{16.72}(1.50)  & \textbf{16.88}(1.50)  & \textbf{17.04}(1.50)  & \textbf{17.18}(1.50)  & \textbf{17.29}(1.50) \\
     &   17.56(2.00)  & 17.85(2.00)  & 18.73(2.00)  & 20.33(2.00) & 22.56(2.00)  & 25.78(2.00)  & 30.46(2.50)  & 37.64(3.00)  & 65.67(3.00)  & 159.89(3.00)\\
     &   17.50(2.00)  & 17.60(1.75)  & 17.57(2.00)  & 17.70(2.00) & 17.70(2.00)  & 17.90(2.00)  & 17.93(1.75)  & 18.03(2.00)  & 18.15(1.75)  & 18.34(1.75) \\
     &   21.39        & 21.39        & 21.46        & 21.55       & 21.67        & 21.83        & 21.99        & 22.25        & 22.57        & 23.13       \\
     &   21.52        & 21.54        & 21.49        & 21.46       & 21.60        & 21.59        & 21.89        & 21.77        & 22.31        & 22.58       \\
     &   21.12        & 25.93        & 38.99        & 59.02       & 86.22        & 119.28       & 156.38       & 196.18       & 237.53       & 281.07      \\
     &   20.92        & 20.92        & 20.95        & 20.99       & 21.04        & 21.12        & 21.22        & 21.38        & 21.67        & 22.14       \\
     \hline
1.75 &   \textbf{9.50}(1.75)  &  \textbf{ 9.69}(1.75)  & 10.23(1.75)  & 11.17(2.00) &  12.64(2.00) &  14.74(2.25) & 17.80(2.75)  &  23.12(3.00) &  43.65(3.00) & 114.15(3.00)\\
     &   \textbf{9.50}(1.75)  &   \textbf{9.53}(1.75)  &  \textbf{9.57}(1.75)  &  \textbf{9.65}(1.75) &   \textbf{9.77}(1.75) &   \textbf{9.91}(1.75) & \textbf{10.09}(1.75)  & \textbf{10.28}(1.75)  &  \textbf{10.53}(1.75) &  10.79(1.75)\\
     &   \textbf{9.51}(1.75)  &   \textbf{9.52}(1.75)  &  \textbf{9.56}(1.75)  &  \textbf{9.64}(1.75) &   \textbf{9.74}(1.75) &   \textbf{9.87}(1.75) & \textbf{10.02}(1.75)  & \textbf{10.20}(1.75)  &  \textbf{10.35}(1.75) & \textbf{10.47}(1.75) \\
     &  10.09(2.25)   & 10.27(2.50)  & 10.79(2.50)  & 11.68(2.50) & 13.09(2.50)  & 15.11(2.50)  & 18.07(3.00)  & 23.34(3.00)  & 43.79(3.00)  & 114.13(3.00)\\
     &  10.07(2.25)   & 10.07(2.50)  & 10.16(2.50)  & 10.17(2.25) & 10.32(2.25)  & 10.41(2.25)  & 10.50(2.25)  & 10.73(2.25)  & 10.84(2.00)  & 11.00(2.25) \\
     &  13.05         & 13.07        & 13.12        & 13.19       & 13.31        & 13.45        & 13.67        & 13.85        & 14.12        & 14.55       \\
     &  11.98         & 11.94        & 11.97        & 11.99       & 12.02        & 12.09        & 12.22        & 12.16        & 12.39        & 12.57       \\
     &  14.93         & 18.28        & 27.53        & 42.02       & 62.30        & 87.99        & 118.23       & 152.47       & 190.09       & 232.36      \\
     &  14.61         & 14.61        & 14.63        & 14.67       & 14.73        & 14.80        & 14.91        & 15.06        & 15.31        & 15.69       \\
     \hline
2.00 &    \textbf{6.60}(2.00)  & 6.70(2.00)    &   7.06(2.00) &  7.67(2.25) &   8.65(2.50) &  10.10(2.75) &  12.29(3.00) &  16.56(3.00) &  32.31(3.00) & 87.30(3.00) \\
     &    \textbf{6.60}(2.00)  & \textbf{6.61}(2.00)   &   \textbf{6.64}(2.00) &  \textbf{6.72}(2.00) &   \textbf{6.82}(2.00) &   \textbf{6.95}(2.00) &   \textbf{7.10}(2.00) &   \textbf{7.30}(2.00) &   \textbf{7.51}(2.00) &  7.71(2.00) \\
     &    \textbf{6.59}(2.00)  & \textbf{6.61}(2.00)   &   \textbf{6.65}(2.00) &  \textbf{6.71}(2.00) &   \textbf{6.80}(2.00) &   \textbf{6.92}(2.00) &   \textbf{7.07}(2.00) &   \textbf{7.24}(2.00) &   \textbf{7.42}(2.00) &  \textbf{7.55}(2.00) \\
     &   6.92(3.00)   & 7.04(3.00)   & 7.36(3.00)   & 7.94(3.00)  & 8.90(3.00)   & 10.32(3.00)  & 12.56(3.00)  & 16.78(3.00)  & 32.46(3.00)  & 87.31(3.00) \\
     &   6.91(2.75)   & 6.93(3.00)   & 6.97(3.00)   & 7.04(2.75)  & 7.12(3.00)   & 7.25(2.75)   & 7.37(3.00)   &  7.55(2.50)  &  7.71(2.50)  &  7.90(2.25) \\
     &   9.19         & 9.21         & 9.25         & 9.33        & 9.44         & 9.58         &  9.75        &  9.94        & 10.19        & 10.54       \\
     &   8.38         & 8.37         & 8.31         & 8.35        & 8.37         & 8.37         & 8.45         & 8.46         & 8.58         & 8.73        \\
     &  11.70         &14.30         & 21.46        & 32.86       & 49.11        & 70.16        & 95.62        & 125.25       & 159.15       & 198.77      \\
     &  11.29         &11.30         &11.32         &11.36        &11.42         &11.49         & 11.60        & 11.74        & 11.96        &  12.28      \\
     \hline

\end{tabular}
\end{sidewaystable}

\begin{sidewaystable}\small
\caption{Out-of-control ARLs of several CUSUM control charts (variance chart for iid observations, LR chart, SPRT chart, Shiryaev-Roberts charts, GLR chart, GSPRT chart and Generalized Shiryaev-Roberts charts) for an in-control ARL of $500$}\begin{tabular}{|c|c|c|c|c|c|c|c|c|c|c|}
\hline
$\Delta/ \phi_1$
     &   0            & 0.1          & 0.2          & 0.3         & 0.4          & 0.5          & 0.6          & 0.7          & 0.8          & 0.9         \\
     \hline
2.25 &     \textbf{5.05}(2.25) &   \textbf{5.13}(2.25) &   5.37(2.50) &  5.80(2.75) &   6.49(3.00) &   7.59(3.00) &   9.39(3.00) &  12.84(3.00) &  25.47(3.00) & 70.02(3.00) \\
     &     \textbf{5.05}(2.25) &   \textbf{5.06}(2.25) &   \textbf{5.09}(2.25) &  \textbf{5.16}(2.25) &   \textbf{5.24}(2.25) &   \textbf{5.37}(2.25) &   \textbf{5.51}(2.25) &   \textbf{5.70}(2.25) &   \textbf{5.90}(2.25) &  6.09(2.25) \\
     &     \textbf{5.04}(2.25) &   \textbf{5.06}(2.25) &   \textbf{5.09}(2.25) &  \textbf{5.15}(2.25) &   \textbf{5.23}(2.25) &   \textbf{5.34}(2.25) &   \textbf{5.47}(2.25) &   \textbf{5.64}(2.25) &   \textbf{5.83}(2.25) &  \textbf{6.00}(2.25) \\
     &   5.28(3.00)   & 5.35(3.00)   & 5.58(3.00)   & 6.02(3.00)  & 6.72(3.00)   & 7.81(3.00)   & 9.59(3.00)   & 13.03(3.00)  & 25.59(3.00)  & 70.00(3.00) \\
     &   5.28(2.75)   & 5.27(3.00)   & 5.31(3.00)   & 5.38(3.00)  & 5.46(3.00)   & 5.58(3.00)   & 5.70(2.75)   & 5.87(3.00)   &  6.05(2.75)  &  6.25(2.75) \\
     &   7.07         & 7.08         & 7.12         & 7.18        & 7.28         & 7.41         & 7.56         & 7.76         &  7.99        &  8.31       \\
     &   6.52         & 6.52         & 6.52         & 6.54        & 6.55         & 6.53         & 6.56         & 6.57         & 6.67         & 6.76        \\
     &   9.67         & 11.80        & 17.72        & 27.15       & 40.74        & 58.60        & 80.54        & 106.66       & 137.20       & 173.97      \\
     &   9.24         & 9.25         & 9.27         & 9.32        & 9.37         & 9.45         & 9.55         & 9.69         &  9.88        & 10.15       \\
     \hline
2.50 &     \textbf{4.12}(2.50) &   \textbf{4.19}(2.50) &   4.36(2.75) &  4.67(3.00) &   5.21(3.00) &   6.10(3.00) &   7.56(3.00) &  10.49(3.00) &  20.94(3.00) & 58.14(3.00) \\
     &     \textbf{4.12}(2.50) &   \textbf{4.13}(2.50) &   \textbf{4.17}(2.50) &  \textbf{4.21}(2.50) &   \textbf{4.29}(2.50) &   \textbf{4.40}(2.50) &   \textbf{4.53}(2.50) &   \textbf{4.71}(2.50) &   \textbf{4.92}(2.50) &  5.12(2.50) \\
     &     \textbf{4.12}(2.50) &   \textbf{4.13}(2.50) &   \textbf{4.16}(2.50) & \textbf{4.21}(2.50)  & \textbf{4.27}(2.50)   & \textbf{4.38}(2.50)   & \textbf{4.51}(2.50)   & \textbf{4.67}(2.50)   & \textbf{4.87}(2.50)   &  \textbf{5.06}(2.50) \\
     &   4.30(3.00)   & 4.36(3.00)   & 4.55(3.00)   & 4.87(3.00)  & 5.43(3.00)   & 6.30(3.00)   & 7.77(3.00)   & 10.65(3.00)  & 21.03(3.00)  & 58.04(3.00) \\
     &   4.30(3.00)   & 4.32(3.00)   & 4.35(3.00)   & 4.37(3.00)  & 4.47(3.00)   & 4.57(3.00)   & 4.70(3.00)   & 4.88(3.00)   & 5.06(3.00)   &  5.25(2.75) \\
     &   5.72         & 5.74         & 5.77         & 5.84        & 5.92         & 6.04         & 6.20         & 6.38         & 6.60         &  6.88       \\
     &   5.44         & 5.45         & 5.44         & 5.45        & 5.47         & 5.46         & 5.49         & 5.49         & 5.55         & 5.63        \\
     &   8.29         & 10.10        & 15.12        & 23.21       & 34.92        & 50.45        & 69.77        & 93.11        & 120.77       & 154.94      \\
     &   7.84         & 7.85         & 7.87         & 7.91        & 7.97         & 8.05         & 8.15         & 8.29         & 8.48         &  8.70       \\
     \hline
2.75 &     \textbf{3.52}(2.75) &   \textbf{3.55}(2.75) &   3.69(3.00) &  3.95(3.00) &   4.38(3.00) &   5.12(3.00) &   6.36(3.00) &   8.87(3.00) &  17.74(3.00) &  49.40(3.00)\\
     &     \textbf{3.52}(2.75) &   \textbf{3.52}(2.75) &   \textbf{3.54}(2.75) &  \textbf{3.59}(2.75) &   \textbf{3.66}(2.75) &   \textbf{3.75}(2.75) &   \textbf{3.88}(2.75) &   \textbf{4.05}(2.75) &   \textbf{4.25}(2.75) &   4.47(2.75)\\
     &   \textbf{3.51}(2.75)   & \textbf{3.52}(2.75)   & \textbf{3.55}(2.75)   & \textbf{3.59}(2.75)  & \textbf{3.66}(2.75)   & \textbf{3.74}(2.75)   & \textbf{3.86}(2.75)   & \textbf{4.02}(2.75)   & \textbf{4.22}(2.75)   & \textbf{4.43}(2.75)  \\
     &   3.67(3.00)   & 3.72(3.00)   & 3.86(3.00)   & 4.13(3.00)  & 4.57(3.00)   & 5.29(3.00)   & 6.53(3.00)   & 9.00(3.00)   & 17.82(3.00)  & 49.43(3.00) \\
     &   3.68(3.00)   & 3.69(3.00)   & 3.71(3.00)   & 3.75(3.00)  & 3.82(3.00)   & 3.92(3.00)   & 4.05(3.00)   & 4.20(3.00)   & 4.40(3.00)   & 4.60(3.00)  \\
     &   4.82         & 4.83         & 4.87         & 4.93        & 5.01         & 5.12         & 5.26         & 5.45         & 5.66         & 5.93        \\
     &   4.76         & 4.76         & 4.76         & 4.76        & 4.75         & 4.77         & 4.79         & 4.79         & 4.83         & 4.89        \\
     &   7.29         & 8.86         & 13.25        & 20.32       & 30.64        & 44.38        & 61.64        & 82.70        & 108.03       & 139.82      \\
     &   6.83         & 6.83         & 6.86         & 6.90        & 6.96         & 7.04         & 7.14         & 7.27         & 7.44         & 7.65        \\
     \hline
3.00 &    \textbf{3.09}(3.00)  &   \textbf{3.12}(3.00) &   3.23(3.00) &  3.44(3.00) &   3.81(3.00) &   4.42(3.00) &   5.50(3.00) &   7.67(3.00) &  15.38(3.00) & 42.89(3.00) \\
     &    \textbf{3.09}(3.00)  &   \textbf{3.10}(3.00) &   \textbf{3.12}(3.00) &  \textbf{3.16}(3.00) &   \textbf{3.22}(3.00) &   \textbf{3.30}(3.00) &   \textbf{3.42}(3.00) &   \textbf{3.57}(3.00) &   \textbf{3.77}(3.00) &   \textbf{4.01}(3.00)\\
     &   \textbf{3.09}(3.00)   & \textbf{3.09}(3.00)   & \textbf{3.12}(3.00)   & \textbf{3.15}(3.00)  & \textbf{3.22}(3.00)   & \textbf{3.30}(3.00)   & \textbf{3.41}(3.00)   & \textbf{3.56}(3.00)   & \textbf{3.76}(3.00)   & \textbf{3.99}(3.00)  \\
     &   3.24(3.00)   & 3.28(3.00)   & 3.38(3.00)   & 3.61(3.00)  & 3.98(3.00)   & 4.59(3.00)   & 5.66(3.00)   & 7.80(3.00)   & 15.46(3.00)  & 42.90(3.00) \\
     &   3.23(3.00)   & 3.25(3.00)   & 3.27(3.00)   & 3.3(3.00)   & 3.38(3.00)   & 3.44(3.00)   & 3.56(3.00)   & 3.72(3.00)   & 3.91(3.00)   & 4.13(2.75)  \\
     &   4.19         & 4.21         & 4.23         & 4.28        & 4.36         & 4.46         & 4.60         & 4.78         & 4.99         & 5.25        \\
     &   4.29         & 4.29         & 4.29         & 4.29        & 4.29         & 4.30         & 4.30         & 4.30         & 4.34         & 4.39        \\
     &   6.54         & 7.92         & 11.82        & 18.11       & 27.33        & 39.68        & 55.31        & 74.52        & 97.81        & 127.50      \\
     &   6.06         & 6.07         & 6.10         & 6.14        & 6.20         & 6.28         & 6.38         & 6.49         & 6.66         & 6.85        \\
     \hline
\end{tabular}
\end{sidewaystable}

\begin{table}
\caption{Average run length (above), worst average delay for $1 \le \tau \le 50$ (middle), and the value of the delay at position $\tau = 50$ (below) for the LR, the SPRT, and the SR chart for optimal reference parameter, and the GLR, the GSPRT chart, and the GSR chart ($\phi_1=0.4$, in-control ARL = 500)}
\vspace*{0.4cm}

\begin{tabular}{|c|c|c|c|c|c|c|}
\hline
    & \textbf{LR} & \textbf{SPRT} & \textbf{SR} & \textbf{GLR} & \textbf{GSPRT} & \textbf{GSR} \\ \hline
$\Delta=1.3$ & 32.52  &  32.59  &  35.09  &  39.40 &  49.18  &  32.68  \\
             & 32.52  &  32.59   &  35.09   &  39.40  &  73.46  &  32.68   \\
             & 29.85 &  29.85  &  30.70  &  33.43 &  73.46  &  21.91  \\
\hline
$\Delta=2.0$ & 7.52      &   6.79  &   7.14  &  9.42  &   8.38  & 11.41  \\
             & 7.52       &   6.79   &   7.14   &  9.42   &   14.22  &  11.41   \\
             & 6.69      &   6.41  &   6.48  &  8.16  &  14.22  &  6.18  \\
\hline
\end{tabular}
\end{table}

\begin{frame}

\begin{figure}[ht!]
\caption{Out-of-control ARLs of several CUSUM charts as a function of the reference value $\Delta^*$ for an in-control ARL of $500$}
\vspace*{0.4cm}

\setlength\tabcolsep{0.8pt}
\begin{tabular}{cc}
\scalebox{0.31}{\includegraphics{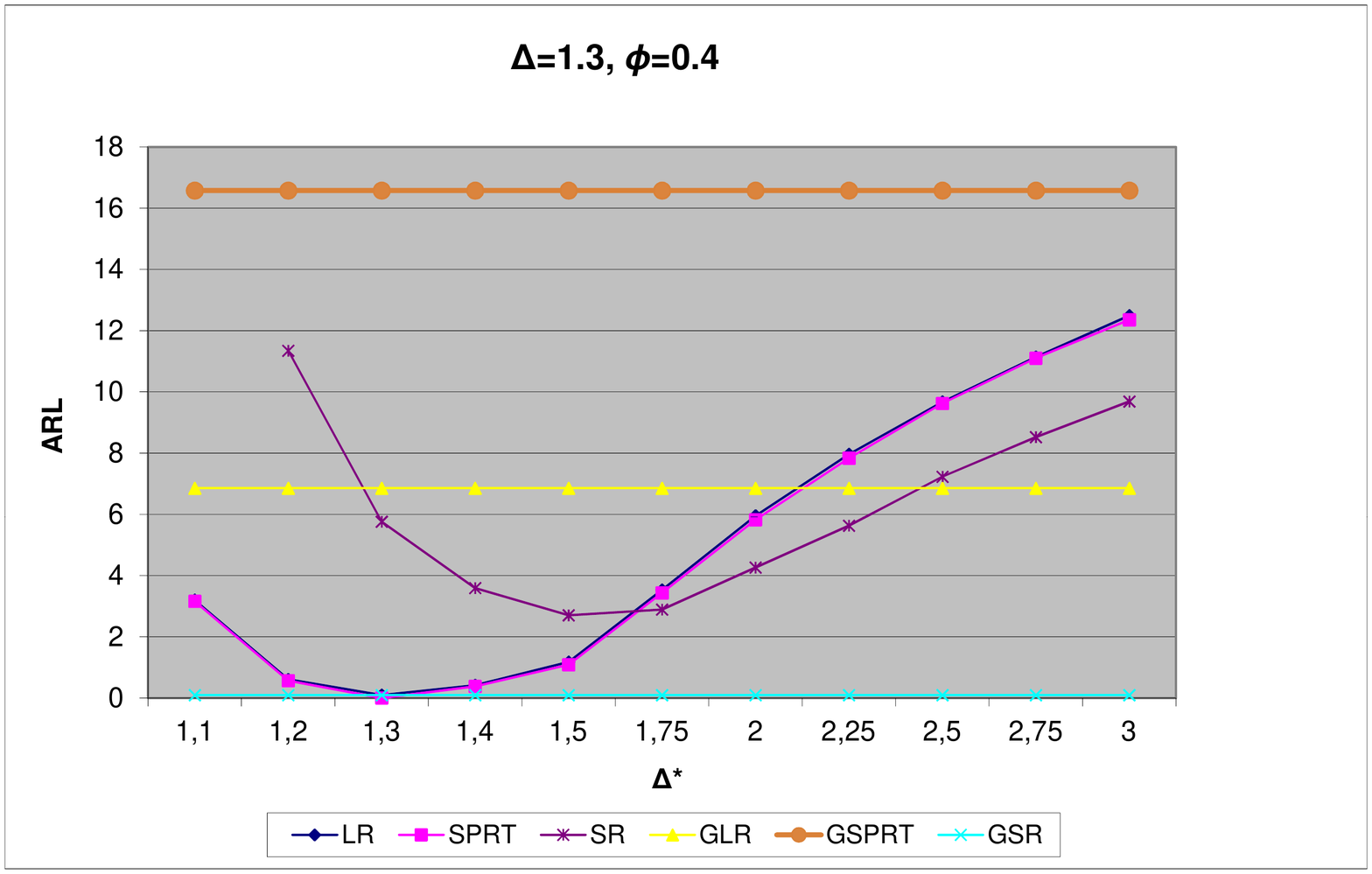}}&
\scalebox{0.31}{\includegraphics{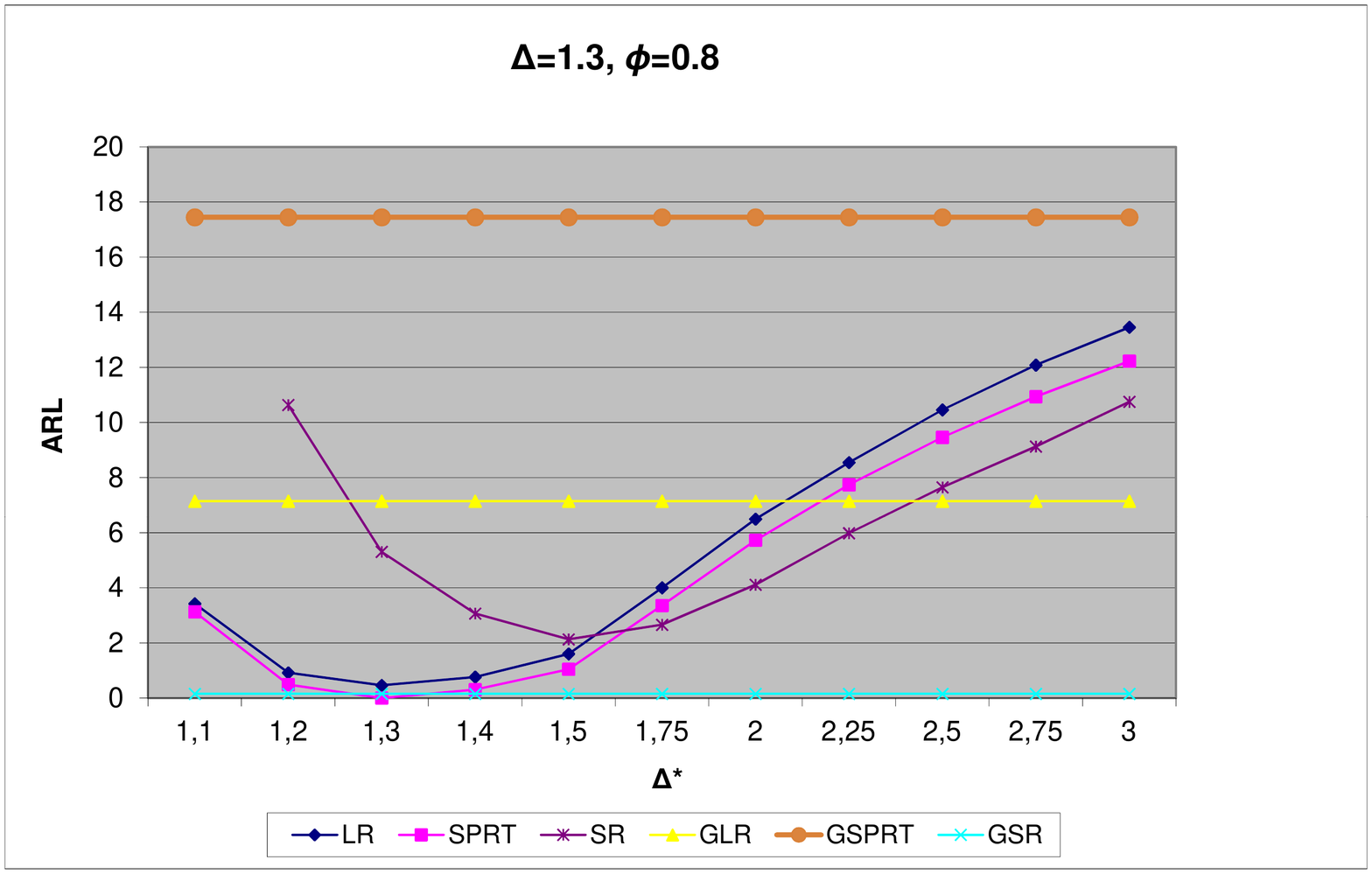}}\\[-0.7cm]
\scalebox{0.31}{\includegraphics{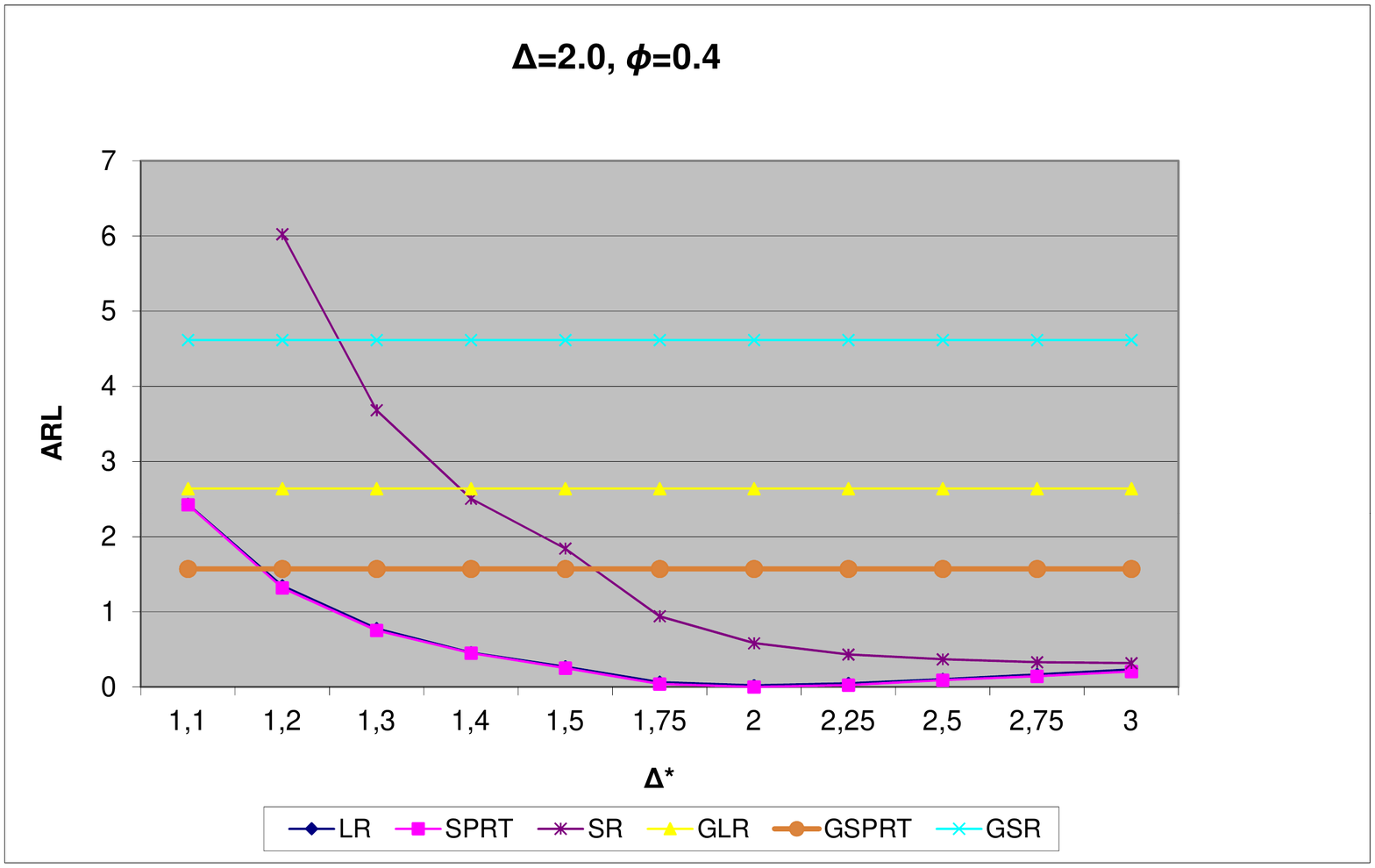}}&
\scalebox{0.31}{\includegraphics{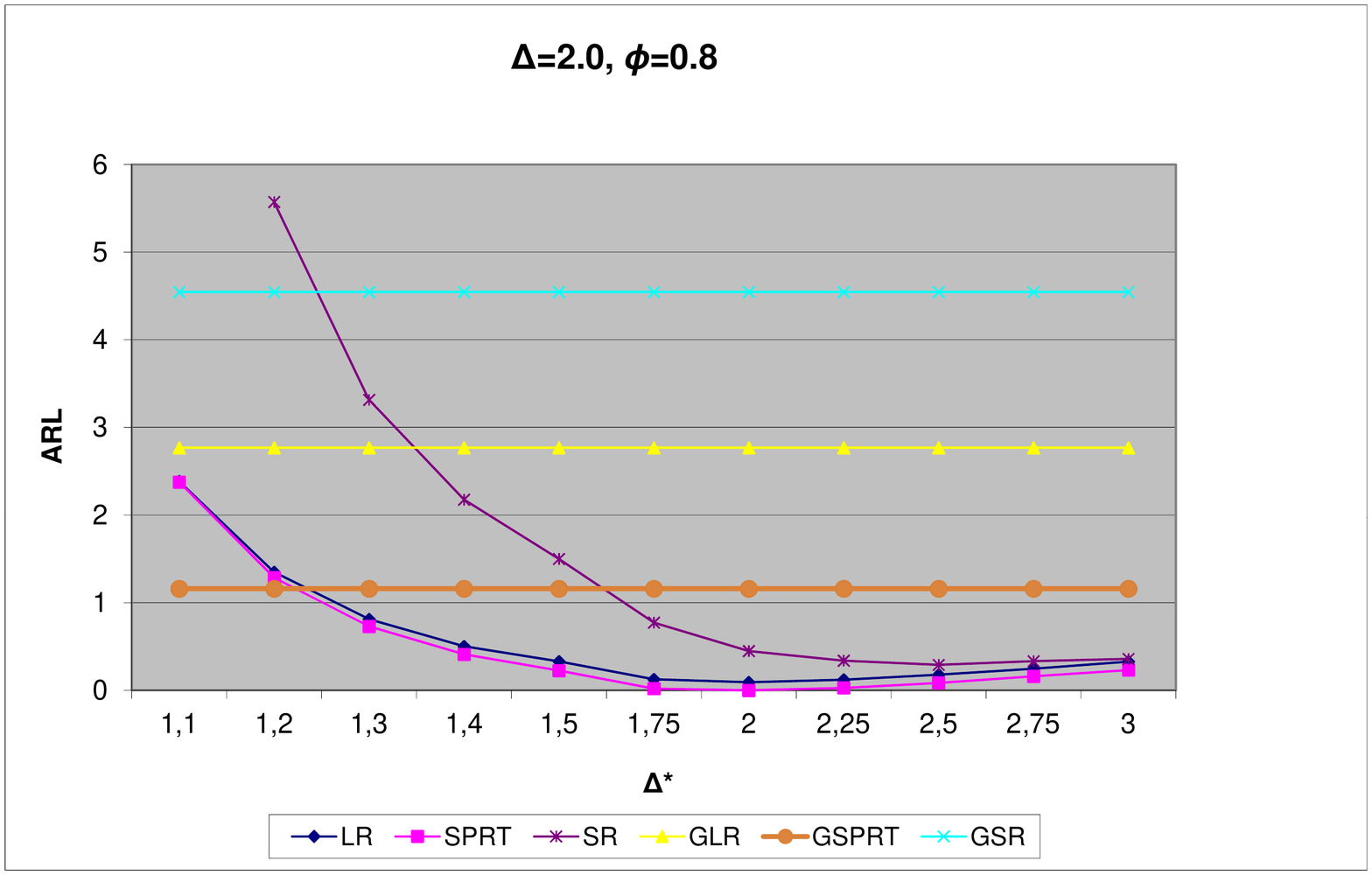}}\\
\end{tabular}
\end{figure}
\end{frame}

\end{document}